\documentclass[twocolumn,prx,amsmath,amssymb,superscriptaddress]{revtex4-1}
\usepackage{graphicx,bm,dsfont,amsmath,amssymb}
\usepackage{soul}
 \usepackage{array}
\usepackage{color}
\usepackage{amsthm}
\usepackage{mathrsfs}
\usepackage{subfigure}
\usepackage{lipsum}
\usepackage{graphicx}
\usepackage{afterpage}
\usepackage{float}
\usepackage[percent]{overpic}
\usepackage[colorlinks=true,citecolor=blue,urlcolor=blue]{hyperref}
\normalsize
\usepackage[ruled,vlined, linesnumbered]{algorithm2e}

\begin{document}

\title{Efficient Preparation of Decoherence Free Subspace Basis States}
\author{Zi-Ming Li}
\affiliation{School of Integrated Circuits, Tsinghua University, Beijing 100084, China}
\author{Yu-xi Liu}\email{yuxiliu@mail.tsinghua.edu.cn}
\affiliation{School of Integrated Circuits, Tsinghua University, Beijing 100084, China}
\affiliation{Frontier Science Center for Quantum Information, Beijing, China}

\date{\today}

\begin{abstract}
Decoherence-free subspace (DFS) provides a crucial mechanism for passive error mitigation in quantum computation by encoding information within symmetry-protected subspaces of the Hilbert space, which are immune from collective decoherence. Constructing a complete set of orthogonal basis states for the DFS is essential to realize fault-tolerant quantum computation by using the DFS codes. However, existing methods for preparing these basis states are often non-scalable, platform-specific, or yield mixed states. Here, we propose a deterministic approach to prepare pure, orthogonal and complete DFS basis states for systems of arbitrary size composed of qubits. Our method employs projective measurements and quantum circuits with single-qubit, two-qubit  and Toffoli gates. We provide a rigorous resource cost analysis both mathematically and numerically. Meanwhile, we demonstrate the realizability of our method on NISQ devices by discussing how to implement our method on a superconducting chip. The proposed method offers a universal solution for preparing the DFS basis states across diverse quantum computing platforms and system sizes, which is realizable in the NISQ era.
\end{abstract}
\maketitle

\section{Introduction}
Decoherence poses one of the most significant obstacles in the realization of scalable quantum information processing. When a qubit, defined by a quantum two-level system or a particle with spin $1/2$, interacts with its environment, coherence between different components of the qubit's wavefunction deteriorates and decoherence occurs. This inevitably leads to errors in quantum information processing. To address such challenge,  quantum error correcting codes, using several physical qubits to encode a single logic qubit, was proposed to protect against various environmental noises~\cite{shor1995scheme,steane1996error,calderbank1997quantum,gottesman1997stabilizer}. Furthermore,  a concept known as the decoherence-free subspace (DFS) has been introduced~\cite{zanardi1997noiseless,lidar1998decoherence} when many qubits interact with a common environment.

A DFS refers to a subspace of the system's Hilbert space wherein the quantum states of  $N$ qubits or spin-$1/2$ particles are unaffected by the decoherence noise. In contrast to quantum error correcting codes, which involve active detection and correction of errors~\cite{shor1995scheme,steane1996error,calderbank1997quantum,gottesman1997stabilizer}, the DFS provides a method of passive error avoidance by exploiting symmetries in the system-environment interaction, thus the quantum coherence and information are preserved without the need for active error correction~\cite{zanardi1997noiseless,lidar1998decoherence}. The states in the DFS  are eigenstates of a total spin zero, which enables the states in the DFS to resist the collective decoherence noise. Since the concept was proposed, the DFS has been implemented in diverse physical platforms~\cite{kielpinski2001decoherence,bourennane2004decoherence,viola1999dynamical,martinis2002rabi}, and explored for realizing fault-tolerant quantum computation~\cite{lidar1998decoherence,kempe2001theory, bacon2000universal}, because the basis states of the DFS can serve as logical qubits for realizing decoherence-free quantum computation. Besides, a special class of quantum states called supersinglets~\cite{cabello2003supersinglets} can be expressed by the basis states of the DFS. Supersinglets are a set of multipartite entangled states with various applications~\cite{cabello2003supersinglets}, characterized by total spin zero~\cite{zanardi1997noiseless} and invariant under collective rotation operations~\cite{zanardi1997noiseless,cabello2002n,cabello2003supersinglets}. Thus, given the basis states of the DFS, one can perform decoherence-free quantum computation or prepare the supersinglets as linear combination of the basis states.

However, some of the DFS basis states  are highly entangled states with complex expressions~\cite{lidar1998decoherence,zanardi1997noiseless, cabello2007six}. Thus, the preparation of the DFS basis states is a highly non-trivial problem, which is one of the main reasons that the DFS has not been widely utilized for realizing fault-tolerant quantum computation. The preparation of the DFS basis states was first proposed and discussed in~\cite{cabello2003supersinglets}.  Experimentally, these states have been prepared via quantum nondemolition measurements~\cite{behbood2013feedback,behbood2014generation} or cavity quantum electrodynamics~\cite{jin2005generation,qiang2011alternative,chen2016fast} on different physical systems. However, the experimental methods are often not universal for different physical platforms and system sizes. Theoretically, methods for preparing the DFS basis states of four and six qubits were proposed~\cite{cabello2003supersinglets,cabello2007six}. A method to prepare supersinglets was also proposed through multiple projection measurements and rotation operators~\cite{ilo2022deterministic}. However, such methods did not provide a universal solution for arbitrary system size or a complexity analysis. Moreover, the final state prepared by Ref.~\cite{ilo2022deterministic} was often a mixed state rather than a pure state, which is more essential for performing quantum computation. Thus, finding a general method for preparing the DFS basis states for different system sizes and physical platforms is an urgent task.

In this work, we propose a theoretical method for deterministically preparing the DFS bases with arbitrary system sizes. The states prepared are all pure states in ideal case. Our method takes the system size as input and outputs a series of quantum states that form a set of orthogonal complete basis states of the DFS. Projective measurements and quantum circuits which consist of single-qubit, two-qubit and Toffoli gates are utilized for realizing our method. The resource cost of the proposed method is calculated and proved both mathematically and numerically. Also, we discuss how to implement our method on real physical platforms with the superconducting chips as an example. We show that the qubit number and running time of the proposed method can be satisfied and realized with NISQ devices. Thus, our approach offers a general solution for preparing the DFS basis states with various system sizes and quantum computing platforms, which is realizable in the NISQ era.

The paper is organized as follows. In Section~\ref{sec2}, we present and explain our method and provide an analysis of its complexity. In Section~\ref{sec3}, we show some examples with numerical calculations to illustrate the reliability of our method. In Sec.~\ref{sec4}, we discuss how to realize the proposed method on real-physical platforms, taking the superconducting quantum computing devices as an example. In Section~\ref{sec5}, we analyze and summarize our results with discussions and conclusions.

\section{Preparation of the DFS Basis States}
\label{sec2}
We explain how to prepare the basis states of the DFS in this section. An introduction of the DFS is given in Appendix.~\ref{ap1}. We consider a physical system consisting of $N$ qubits. We mainly provide a method to find a set of orthogonal complete bases of the DFS. It is noted that the DFS is a subspace where all states are eigenstates of  a total spin zero, thus  the number $N$  of the total qubits is even.

We first show how to find a complete but not orthogonal set of bases for the DFS. Given the value of $N$, it is proved that the dimension of the DFS is $d(N)$ with Ref.~\cite{zanardi1997noiseless,cabello2007six}
\begin{equation}
	d(N)=\frac{N!}{(N/2)!(N/2+1)!}.
\end{equation}
When $N=2$, the value of $d(N)$ is $1$, thus there is only one state in the DFS for $N=2$
\begin{equation}
	|\psi\rangle=\frac{1}{\sqrt{2}}\left(|01\rangle-|10\rangle\right).
\end{equation}
$|\psi\rangle$ is also known as the singlet state for a pair of qubits.

When $N>2$, we can split the whole system with $N$ qubits into $N/2$ subsystems where each subsystem contains two qubits. Suppose each of the subsystems is in the singlet state $|\psi\rangle$, then the state of the whole system can be written as tensor products of $N/2$ subsystems as
\begin{equation}
	\label{eq1}
	|\phi\rangle=|\psi_{i_1i_2}\rangle\otimes |\psi_{i_3i_4}\rangle \otimes \cdots \otimes |\psi_{i_{N-1}i_N}\rangle,
\end{equation}
where $\{i_1,i_2,\cdots,i_N\}$ is a permutation of $\{1,2,\cdots,N\}$ and each $|\psi_{i_mi_{m+1}}\rangle$ denotes a subsystem in the singlet state which contains the $i_m$th and the $i_{m+1}$th qubit. Thus, the state of the whole system $|\phi\rangle$ has a total spin zero and is in the DFS. By permuting the order of $\{i_1,i_2,\cdots,i_N\}$, we can get a series of states with total spin zero in the DFS. The number $f(N)$ of different quantum states, that can be constructed by permutation the order of $\{i_1,i_2,\cdots,i_N\}$ in Eq.~(\ref{eq1}),  is
\begin{equation}
	f(N)=(N-1)!!=\frac{N!}{2^{N/2} (N/2)!}>d(N).
\end{equation}
By choosing $d(N)$ linear independent states from the $f(N)$ states, a set of complete but not orthogonal bases can be constructed. For example, when $N=6$, we find $d(N)=5$ and $f(N)=30$. We can choose five linear independent states
\begin{align}
	\label{eq7}
	|a_1\rangle &= |\psi_{12}\rangle \otimes|\psi_{34}\rangle\otimes|\psi_{56}\rangle, \nonumber\\
	|a_2\rangle &= |\psi_{12}\rangle \otimes|\psi_{35}\rangle\otimes|\psi_{46}\rangle, \nonumber\\
	|a_3\rangle &= |\psi_{13}\rangle \otimes|\psi_{24}\rangle\otimes|\psi_{56}\rangle,\nonumber\\
	|a_4\rangle &= |\psi_{13}\rangle \otimes|\psi_{25}\rangle\otimes|\psi_{46}\rangle, \nonumber\\
	|a_5\rangle &= |\psi_{14}\rangle \otimes|\psi_{25}\rangle\otimes|\psi_{36}\rangle,
\end{align}
as a set $\{|a_1\rangle,|a_2\rangle,|a_3\rangle,|a_4\rangle,|a_5\rangle\}$  of complete bases. The method for choosing $d(N)$ linear independent states out of the $f(N)$ states is detailed in Appendix.~\ref{ap2}. Thus, for each $N$, we can construct a set of complete bases $\{|a_1\rangle, |a_2\rangle,\cdots,|a_{d(N)}\rangle\}$ for the DFS.
 \begin{figure}[t]
	\includegraphics[width=\linewidth]{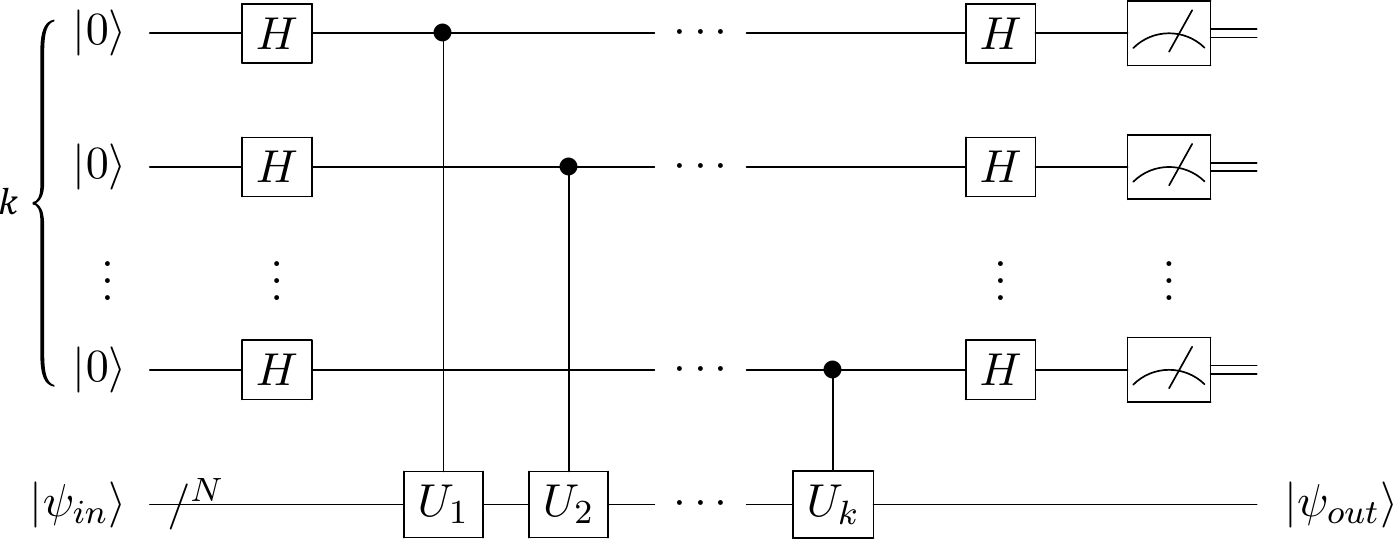}
	\caption{The quantum circuit for preparing state $|u_{k+1}\rangle$ based on $\{|a_1\rangle,|a_2\rangle,\cdots,|a_{k+1}\rangle\}$. The circuit contains two quantum registers. The first register contains the top $k$ qubits as ancliiary qubits, initialized in $|0\rangle^{\otimes k}$. And the second register contains the bottom $N$ qubits as system qubits, initialized in $|\psi_{in}\rangle$. The $H$ gate in the circuit is the Hadamard gate, and each $U_i$ is taken as $U_i=I-2|a_i\rangle\langle a_i|$,$\forall i=1,2,\cdots,k$. All $k$ ancillary qubits are measured at then end of the quantum circuit in the computational bases.}
	\label{fig1}
\end{figure}

Following the Gram-Schmidt process, we can define a set  $\{|t_1\rangle,|t_2\rangle,\cdots,|t_{d(N)}\rangle\}$ of states as that of orthogonal complete bases for the DFS by using the complete set of states $\{|a_1\rangle, |a_2\rangle,\cdots,|a_{d(N)}\rangle\}$. Each state $|t_{k+1}\rangle$ is defined as
\begin{equation}
	\label{eq4}
	|t_{k+1}\rangle =\frac{|a_{k+1}\rangle-\mathcal{P}_k|a_{k+1}\rangle}{\parallel |a_{k+1}\rangle-\mathcal{P}_k|a_{k+1}\rangle\parallel},
\end{equation}
where $\mathcal{P}_k$ is the projector onto the subspace $span\{|a_1\rangle,|a_2\rangle,\cdots,|a_k\rangle\}$, $ \forall k=0,1,\cdots,d(N)-1$. Thus, each $|t_{k+1}\rangle$ satisfies $|t_{k+1}\rangle\in span\{|a_1\rangle,|a_2\rangle,\cdots,|a_{k+1}\rangle\}$ and can be written as a linear combination of $\{|a_1\rangle,|a_2\rangle,\cdots,|a_{k+1}\rangle\}$. We now show how to prepare a state that approaches the state $|t_{k+1}\rangle$ with arbitrary accuracy based on $\{|a_1\rangle,|a_2\rangle,\cdots,|a_{k+1}\rangle\}$, $\forall k=0,1,\cdots,d(N)-1$. We denote the prepared set of states to be $\{|u_1\rangle,|u_2\rangle,\cdots,|u_{d(N)}\rangle\}$, and it is required that for a given $\epsilon>0$,
\begin{equation}
	1-\parallel \langle u_i|t_i\rangle\parallel^2<\epsilon, \forall i=1,2,\cdots,d(N).
\end{equation}

\begin{figure}[t]
		\includegraphics[width=.6\linewidth]{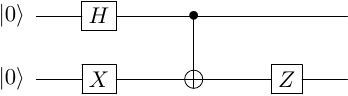}
		\caption{The quantum circuit for preparing the two-qubit singlet state $|\psi\rangle=(|01\rangle)-|10\rangle)/\sqrt{2}$. The quautm state as in Eq.~(\ref{eq1}), i.e, each $|a_i\rangle$ that we defined $\forall i=1,\cdots,d(N)$, can be implemented with this circuit. The quautm state as in Eq.~(\ref{eq1}) is prepared by applying this circuit on each subsystem labelled by $|\psi_{i_mi_{m+1}}\rangle$, respectively. Thus, $O_{a_i}$ and $U_i$ defined in Eq.~(\ref{eq2}) and Eq.~(\ref{eq3}) can be implemented with constant number of singlet-qubit gates and controlled-NOT gates.}
				\label{fig2}
\end{figure}

The preparation of $|u_{k+1}\rangle$ to approximate $|t_{k+1}\rangle$ can be realized by the quantum circuit as shown in Fig.~\ref{fig1}, with each $U_i$ taken as $I-2|a_i\rangle\langle a_i|$, $\forall i=1,2,\cdots,k$. There are two quantum registers in this quantum circuit: the ancillary register containing the first $k$ qubits initialized in $|0\rangle^{\otimes k}$ and the system register containing $N$ qubits. The quantum circuit in Fig.~\ref{fig1} can be implemented efficiently via single-qubit, two-qubit and Toffoli gates. We assume that each $|a_i\rangle$ can be prepared with $O_{a_i}$ as
\begin{equation}
	\label{eq2}
	O_{a_i}|0\rangle=|a_i\rangle, \forall i=1,2,\cdots,d(N),
\end{equation}
then $U_i$ in the circuit can be implemented through
\begin{equation}
	\label{eq3}
	U_i=O_{a_i}(I-2|0\rangle\langle 0|)O_{a_i}^{\dagger},\forall i=1,2,\cdots,d(N).
\end{equation}
As given in Eq.~(\ref{eq1}),  each $|a_i\rangle$ has a form of tensor product of a series singlet states, which can be  implemented with quantum circuits using certain number of single-qubit gates and controlled-NOT gates as shown in Fig.~\ref{fig2}. Each $U_i$ can be implemented with certain number of single qubit gates and controlled-NOT gate, which can be realized efficiently on physical system without using any quantum oracles. Thus, the whole quantum circuit as in Fig.~\ref{fig1} can be realized efficiently with only single-qubit gates, two-qubit gates, and Toffoli gates with no quantum oracles.

Suppose the input of the second register of the quantum circuit is $|\psi_{in}\rangle$, which is an arbitrary state. Then the output state of the quantum circuit is
\begin{equation}
	\label{eq12}
	\sum_{q_1=0}^1\sum_{q_2=0}^1\cdots \sum_{q_k=0}^1 \prod_{i=k}^{1}|q_1q_2\cdots q_k\rangle \frac{I+(-1)^{q_i}U_i}{2}|\psi_{in}\rangle.
\end{equation}
Suppose when we measure all the $k$ ancillary qubits at the end of the quantum circuit, the $k$ outcomes of the measurements are all $0$, then the final state $|\psi_{out}\rangle$ of the system register is
\begin{align}
	|\psi_{out}\rangle&\propto\prod_{i=k}^{1}\frac{I+U_i}{2}|\psi_{in}\rangle\nonumber\\
	&=(I-|a_k\rangle\langle a_k|)\cdots (I-|a_1\rangle\langle a_1|)|\psi_{in}\rangle.\nonumber\\
\end{align}
If we denote $P_i=I-|a_i\rangle\langle a_i|$ as the projector onto the subspace orthogonal to $|a_i\rangle$, and we define
\begin{equation}
	Q_i=P_iP_{i-1}\cdots P_1, , \forall i=1,2,\cdots,d(N)
\end{equation}
Then the relationship between $|\psi_{out}\rangle$ and $|\psi_{in}\rangle$  is
\begin{equation}
	\label{eq18}
	|\psi_{out}\rangle=\frac{P_kP_{k-1}\cdots P_1|\psi_{in}\rangle}{\parallel P_kP_{k-1}\cdots P_1|\psi_{in}\rangle \parallel}=\frac{Q_k|\psi_{in}\rangle}{\parallel Q_k|\psi_{in}\rangle\parallel},
\end{equation}
when the $k$ outcomes of the measurements in ancillary register are all $0$.

\begin{figure*}[th]
	\centering
	\subfigure {\
		\begin{minipage}[h]{0.18\linewidth}
			\centering
			\begin{overpic}[scale=0.43]{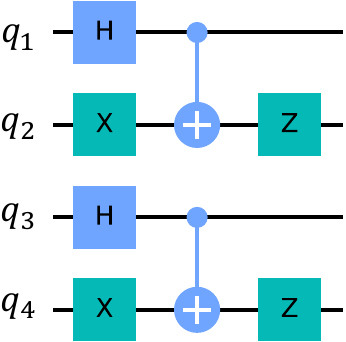}
				\put(0,110){\textbf{(a)}}
			\end{overpic}
		\end{minipage}
	\label{fig3}
	}
	\subfigure {\
		\begin{minipage}[h]{0.75\linewidth}
			\centering
			\begin{overpic}[scale=0.43]{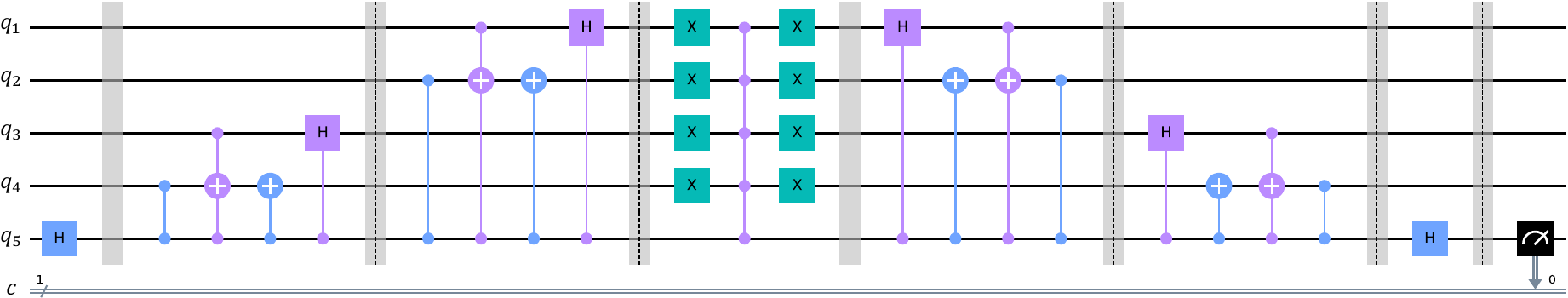}
				\put(0,21.7){\textbf{(b)}}
			\end{overpic}
		\end{minipage}
	\label{fig4}
	}
	\caption{(a)The quantum circuit for preparing the state $|u_1\rangle$ for the $N=4$ case. All qubits are initialized in $|0\rangle$. (b)The quantum circuit for preparing the state $|u_2\rangle$ for the $N=4$ case. The qubit $q_5$ is an ancillary qubit which needs to be measured as in Fig.~\ref{fig1}. The qubits $q_1$ to $q_4$ are the system qubits for storing $|u_2\rangle$. The qubit $q_5$ is initialized in $|0\rangle$, while the qubits $q_1$ to $q_4$ are initialized in $|a_2\rangle$ as defined in Eq.~(\ref{eq8}). The black dashed lines are barrires which are drawn to enhance the readability of the quantum circuit.}
\end{figure*}

Following the quantum circuit in Fig.~\ref{fig1}  , we now describe how to construct the state $|u_{k+1}\rangle$ to approach the state $|t_{k+1}\rangle$ as in Eq.~(\ref{eq4}) with arbitrary accuracy $\epsilon$. In the first step, we initialize the input of the second register of the quantum circuit as in Fig.~\ref{fig1} to be $|a_{k+1}\rangle$, and  all qubits in the first register are initialized to the state $|0\rangle$,  then we run the quantum circuit and measure all $k$ ancillary qubits in the computational bases. If all measurement outcomes are $0$, then the state in the second register is proportional to $Q_k|a_{k+1}\rangle$ following Eq.~(\ref{eq18}). If at least one of $k$ measurement outcomes is nonzero, then the ouput state is abandoned, and we start over the construction procedure till all measurement outcomes are $0$. Thus we can construct a state proportional to $Q_k|a_{k+1}\rangle$, which is stored in the second register.

In the second step,  using the state $Q_k|a_{k+1}\rangle$ obtained in the first step as the input for the second register,  and initializing all $k$ qubits of the first register in the state $|0\rangle$, we run the quantum circuit as shown in Fig.~\ref{fig1} again, and perform measurements of all $k$ ancillary qubits in the first register. If all measurement outcomes are $0$, then the state in the second register is proportional to $(Q_k)^2|a_{k+1}\rangle$, else the output state is abandoned, and we start over the construction proceduce, going back to the first step where the input state of the system register is initialized in $|a_{k+1}\rangle$ till we successfully construct a state proportional to $(Q_k)^2|a_{k+1}\rangle$.

Following this procedure, we can successively construct state that is proportional to $(Q_k)^3|a_{k+1}\rangle$, $(Q_k)^4|a_{k+1}\rangle$,$\cdots$,$(Q_k)^{m_{k+1}}|a_{k+1}\rangle, \forall m_{k+1}\in \mathbb{Z}_+$, where the subscript $k+1$ in $m$ denotes the $(k+1)$th state. That is, $m_{k+1}$ means that the operator $Q_k$ acts on the $(k+1)$th state $|a_{k+1}\rangle$ with $m_{k+1}$ times. We denote $|a_{k+1}^{(m_{k+1})}\rangle$ to be
\begin{equation}
	\label{eq11}
	|a_{k+1}^{(m_{k+1})}\rangle=\frac{(Q_k)^{m_{k+1}}|a_{k+1}\rangle}{\parallel (Q_k)^{m_{k+1}}|a_{k+1}\rangle \parallel}\propto(Q_k)^{m_{k+1}}|a_{k+1}\rangle.
\end{equation}
When $m_{k+1}$ is large enough, we denote $|a_{k+1}^{(m_{k+1})}\rangle$ as $|u_{k+1}\rangle$, which can approximate the expected ideal state $|t_{k+1}\rangle$ with the accuracy $\epsilon$. This is because when $m_{k+1}$ tends to infinity, we have
\begin{equation}
	\label{eq5}
	\lim\limits_{m_{k+1}\to +\infty} (Q_k)^{m_{k+1}}=I-\mathcal{P}_k,
\end{equation}
with $\mathcal{P}_k$ defined in Eq.~(\ref{eq4}) as the projector onto the subspace $span\{|a_1\rangle,|a_2\rangle,\cdots,|a_k\rangle\}$. Thus, we have
\begin{align}
	\lim\limits_{m_{k+1}\to +\infty} |u_{k+1}\rangle &=\lim\limits_{m_{k+1}\to +\infty} |a_{k+1}^{(m_{k+1})}\rangle\nonumber\\
	&=\lim\limits_{m_{k+1}\to +\infty}\frac{(Q_k)^{m_{k+1}}|a_{k+1}\rangle}{\parallel (Q_k)^{m_{k+1}}|a_{k+1}\rangle \parallel}\nonumber\\
	&=\frac{|a_{k+1}\rangle-\mathcal{P}_k|a_{k+1}\rangle}{\parallel |a_{k+1}\rangle-\mathcal{P}_k|a_{k+1}\rangle\parallel}=|t_{k+1}\rangle
\end{align}
The proof of Eq.~(\ref{eq5}) is given in Appendix.~\ref{ap3}.

Therefore, we can denote the state $|a_{k+1}^{(m_{k+1})}\rangle$ to be the constructed $|u_{k+1}\rangle$ for some large enough $m_{k+1}$. That is, we can construct the state $|u_{k+1}\rangle$ to approximate $|t_{k+1}\rangle$, $\forall k=0,1,\cdots,d(N)-1$. Following this procedure, we can construct a set of orthogonal complete bases $\{|u_1\rangle, |u_2\rangle,\cdots,|u_{d(N)}\rangle\}$ for the DFS, accomplishing the task of preparation the basis states of the DFS. The resource cost for our preparation procedure, i.e., the number of qubits and quantum gates that are needed, is described as follows.

We define a matrix $A_{d(N)}=\left(|a_1\rangle,|a_2\rangle,\cdots,|a_{d(N)}\rangle\right)$, and $\kappa$ as the conditional number of the matrix $A_{d(N)}$, i.e., the ratio of the maximum singular value $\sigma_{\max}$ and the minimum singular value $\sigma_{\min}$. Then we can prepare a set  $\{|u_1\rangle, |u_2\rangle,\cdots,|u_{d(N)}\rangle\}$ of basis states for the DFS consisting of $N$ qubits with the methods that we propose, using a total number of $O(d(N))$ qubits and
\begin{equation}
	\label{eq10}
	O\left(d(N)^2\kappa^2\ln\left(\frac{\kappa}{\epsilon}\right)\right)
\end{equation}
quantum gates, such that $\forall k=0, 1,\cdots,d(N)-1$ and $\epsilon>0$, it satisfies the condition
\begin{equation}
	\label{eq6}
	1-\parallel \langle u_{k+1}| t_{k+1}\rangle \parallel^2<\epsilon.
\end{equation}
Meanwhile, the value of each $m_{k+1}$ is
\begin{equation}
	m_{k+1}=O\left(\ln\left(\frac{\kappa}{\epsilon}\right)\right)
\end{equation}
so that $|a_{k+1}^{(m_{k+1})}\rangle$ can be denoted as $|u_{k+1}\rangle$.
The proof of the results described above is given in Appendix.~\ref{ap4}.

\section{Numerical Validations}
\label{sec3}
We now carry out the numerical simulation of the proposed quantum algorithm for validation. For the case $N=4$ and $N=6$, we construct quantum circuits with single-qubit, two-qubit and multi-qubit Toffoli gates, respectively. For the $N=4$ case, $d(N=4)=2$ and thus we choose the set of complete bases to be $\{|a_1\rangle, |a_2\rangle\}$, where
\begin{align}
	\label{eq8}
	|a_1\rangle &= |\psi_{12}\rangle \otimes|\psi_{34}\rangle \nonumber\\
	&= \frac{1}{2}(|0101\rangle-|0110\rangle-|1001\rangle+|1010\rangle),\nonumber\\
	|a_2\rangle &= |\psi_{13}\rangle \otimes|\psi_{24}\rangle \nonumber\\
	&=\frac{1}{2}(|0011\rangle-|0110\rangle-|1001\rangle+|1100\rangle),
\end{align}
to prepare $\{|t_1\rangle, |t_2\rangle\}$, where
\begin{align}
	|t_1\rangle = |a_1\rangle= \frac{1}{2}(|0101\rangle-|0110\rangle-|1001\rangle+|1010\rangle),\nonumber
\end{align}
\begin{align}
	|t_2\rangle=&\frac{1}{2\sqrt{3}}(2|0011\rangle-|0110\rangle-|1001\rangle\nonumber\\
	&-|0101\rangle-|1010\rangle+2|1100\rangle).
\end{align}

 We show the constructed quantum circuit for preparing the quantum states $|u_1\rangle$ and $|u_2\rangle$ for the case of $N=4$ in Fig.~\ref{fig3} and Fig.~\ref{fig4}, respectively. We note that the circuit in  Fig.~\ref{fig4} for preparing the state $|u_2\rangle$ includes one ancillary qubit denoted by $q_0$.  All the quantum circuits are drawn with Qiskit, a Python package for quantum computation developed by IBM~\cite{Qiskit2021}.


 For the $N=6$ case, $d(N=6)=5$, the set $\{|a_1\rangle,|a_2\rangle,|a_3\rangle,|a_4\rangle,|a_5\rangle\}$ of complete basis states is chosen as in Eq.~(\ref{eq7}).
\begin{figure}[b]
	\includegraphics[width=\linewidth]{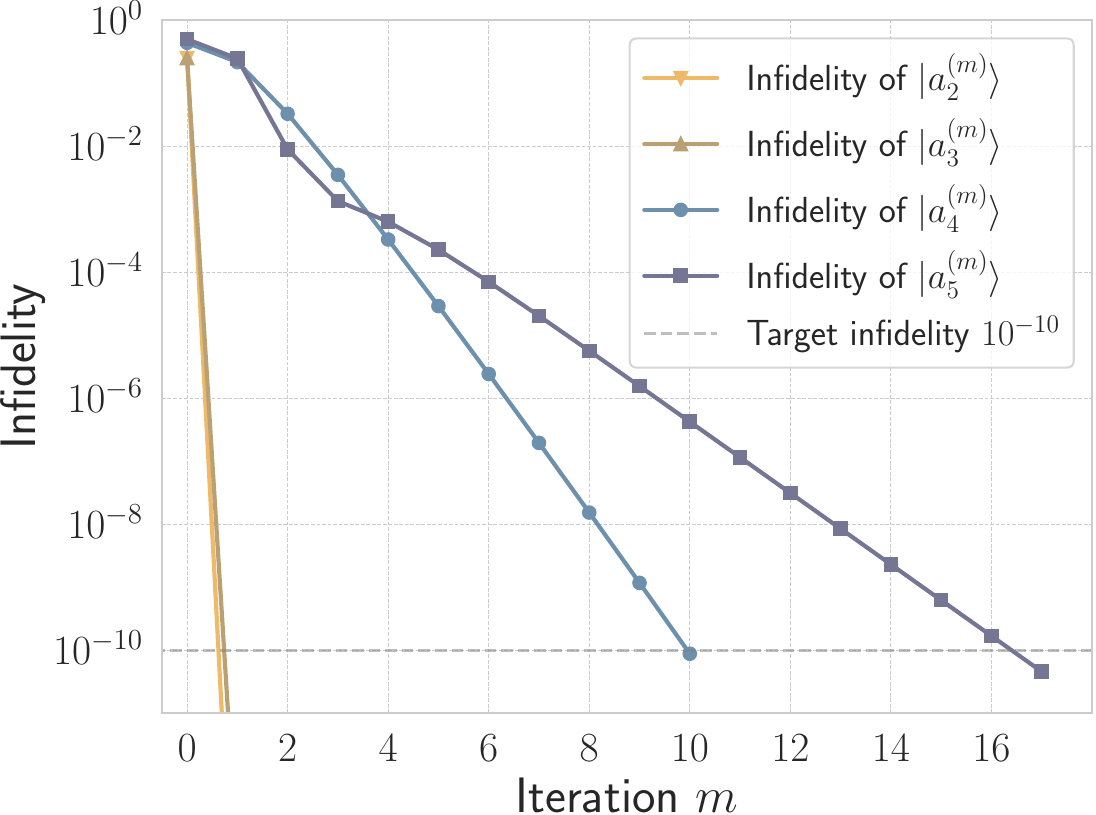}
	\caption{The relationship between the infidelity of $|a_k^{(m_k)}\rangle$ and iteration $k$ for the $N=6$ case. The infidelity decreases exponentially with $m_k$ for all $k=2,3,\cdots,d(N)$.}
	\label{fig5}
\end{figure}
We run the constructed quantum circuit as shown in  Appendix.~\ref{ap6}  to prepare all five basis states in the set $\{|a_1\rangle,|a_2\rangle,|a_3\rangle,|a_4\rangle,|a_5\rangle\}$ of orthogonal complete states. The circuits run with Qiskit by a noiseless quantum simulator. We first calculate how large the value of $m_{k+1}$ needs to be such that the infidelity
\begin{figure}[b]
	\includegraphics[width=\linewidth]{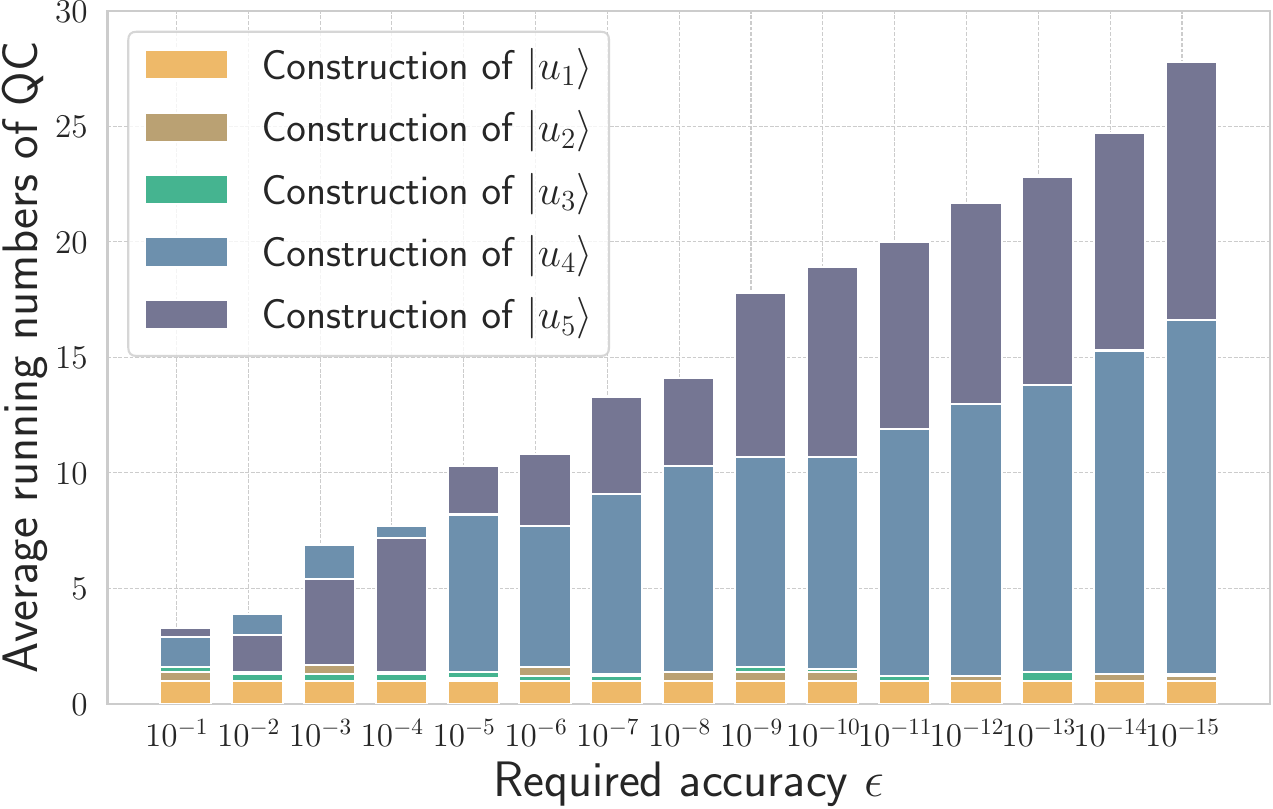}
	\caption{The relationship between the average running number of quantum circuit as defined in Fig.~\ref{fig1} for constructing each $|u_k\rangle,\forall k=1,2,\cdots,d(N)$ and the required infedility $\epsilon$. For the $N=6$ case, the results are shown. The average number of runs of quantum circuits needed for constructing $|u_k\rangle$ grows only linearly with the value of $\ln(1/\epsilon)$, $\forall k=1,2,\cdots,d(N)$.}
	\label{fig6}
\end{figure}
\begin{equation}
	\label{eq9}
		1-\parallel \langle a_{k+1}^{(m_{k+1})}|\frac{|a_{k+1}\rangle-\mathcal{P}_k|a_{k+1}\rangle}{\parallel |a_{k+1}\rangle-\mathcal{P}_k|a_{k+1}\rangle\parallel} \parallel^2<\epsilon
\end{equation}
is satisfied for the state $|a_{k+1}^{(m_{k+1})}\rangle$, which can be used to denote the state $|u_{k+1}\rangle$, $\forall k=0,1,\cdots,4$. We examine the preformance of the algorithm and show the results for the $N=6$ case in Fig.~\ref{fig5}. In the $N=6$ case, $d(N)=5$, and as we take $|u_1\rangle=|a_1\rangle$, we only need to construct $\{|a_2\rangle,|a_3\rangle,|a_4\rangle,|a_5\rangle\}$. It is shown from Fig.~\ref{fig5} that the state $|a_2^{(1)}\rangle$ and $|a_3^{(1)}\rangle$ give a good approximation of the target state immediately with the infidelity much less than the target infidelity, which is set to be $10^{-10}$. And the infidelity of state $|a_4^{(m_{4})}\rangle$ and $|a_5^{(m_5)}\rangle$ decreases exponentially with $m_4$ and $m_5$. The results show the exponential convergence of $|a_k^{(m_k)}\rangle$ to the target state, $\forall k=1,2,\cdots,5$, which indicates the correctness and efficiency of the proposed algorithm. Thus the target orthogonal set of states $\{|t_1\rangle,|t_2\rangle,|t_3\rangle,|t_4\rangle,|t_5\rangle\}$ is constructed, which is shown in Eq.~(\ref{eq14}).  It is noted that due to the unitary degree of freedom for choosing the bases of a linear space, the set of $\{|t_1\rangle, \cdots,|t_{d(N)}\rangle\}$ does not necessarily equal to the states chosen in our algorithm.

\begin{figure*}[!t]
	\normalsize
	\setcounter{equation}{22}
	\begin{align}
		|t_1\rangle=\frac{1}{2\sqrt{2}}(|010101\rangle-|010110\rangle-|011001\rangle+|011010\rangle-|100101\rangle+|100110\rangle+|101001\rangle-|101010\rangle), \nonumber
		\end{align}
	\begin{align}
	|t_2\rangle=\frac{1}{2\sqrt{6}}&(2|010011\rangle-|010101\rangle-|010110\rangle-|011001\rangle-|011010\rangle+2|011100\rangle\nonumber\\
	&-2|100011\rangle+|100101\rangle+|100110\rangle+|101001\rangle+|101010\rangle-2|101100\rangle),\nonumber
	\end{align}
	\begin{align}
		|t_3\rangle=\frac{1}{2\sqrt{6}}&(2|001101\rangle-2|001110\rangle-|010101\rangle+|010110\rangle-|011001\rangle+|011010\rangle\nonumber\\
		&-|100101\rangle+|100110\rangle-|101001\rangle+|101010\rangle+2|110001\rangle-2|110010\rangle),\nonumber
	\end{align}
	\begin{align}
		|t_4\rangle=\frac{1}{6\sqrt{2}}&(4|001011\rangle-2|001101\rangle-2|001110\rangle-2|010011\rangle+|010101\rangle+|010110\rangle-|011001\rangle-|011010\rangle+2|011100\rangle\nonumber\\
		&-2|100011\rangle+|100101\rangle+|100110\rangle-|101001\rangle-|101010\rangle+2|101100\rangle+2|110001\rangle+2|110010\rangle-4|110100\rangle),\nonumber
	\end{align}
\begin{align}
	\label{eq14}
	|t_5\rangle=\frac{1}{6}&(3|000111\rangle-|001011\rangle-|001101\rangle-|001110\rangle-|010011\rangle-|010101\rangle-|010110\rangle+|011001\rangle+|011010\rangle+|011100\rangle\nonumber\\
	&-|100011\rangle-|100101\rangle-|100110\rangle+|101001\rangle+|101010\rangle+|101100\rangle+|110001\rangle+|110010\rangle+|110100\rangle-3|111000\rangle).
\end{align}
	\hrulefill
	\vspace*{2pt}
\end{figure*}

We now examine the complexity of the algorithm with the choice of different $\epsilon$. By taking into account all iterations and the cases where the measurement outcomes of the ancillary qubits are not all zero, we count the number that quantum circuits as in Fig.~\ref{fig1} need to be run to construct each $|u_k\rangle$ from the initial state $|a_k\rangle$. We show the results of the $N=6$ case in Fig.~\ref{fig6}. It is shown that the average number of runs of quantum circuits needed for constructing $|u_k\rangle$ grows linearly with the value of $\ln(1/\epsilon)$, $\forall\, k=1,2,\cdots,5$. This result is consistent with Eq.~(\ref{eq10}) for  $N$ qubits, demonstrating correctness and efficiency of the proposed algorithm and validating the main result of the paper.

\section{Physical realization with quantum chip of superconducting qubits}
\label{sec4}
Our method can be applied to many physical systems. However, for concreteness,  we here use an example of the superconducting qubit circuits to further show how to prepare the set of orthogonal complete basis states for the DFS. We assume that all qubits have switchable nearest-neighbor XY interactions. We consider an $N$-qubit system with Hamiltonian
\begin{equation}
	\label{eq13}
	H=\frac{\omega}{2}\sum_{i=1}^{N}\sigma_i^z+\sum_{\langle i,j\rangle} g_{ij}(\sigma_i^x\sigma_j^x+\sigma_i^y\sigma_j^y),
\end{equation}
where $\sigma^x$, $\sigma^y$, $\sigma^z$ are Pauli operators. Each qubit has the same frequency $\omega$, the summation over $\langle i,j\rangle$ denotes a pair of neighbouring qubits with coupling strength $g_{ij}$. It is noted that in this section we take $\hbar=1$, and the coupling  strength $g_{ij}$ between neighbouring qubits can be switched on and off. This is a commonly used model in superconducting quantum circuits. For instance, each qubit can be a transmon, and the qubits interact with each other through tunable couplers~\cite{gu2017microwave,stehlik2021tunable,ye2021realization}.

By switching off  the couplings between a target qubit from all of its neighboring qubits, one can apply a pulse to the target qubit or let the target qubit  evolve freely for implementing single-qubit rotation gates
\begin{equation}
	R_{\mathrm{X}}(\theta)=\exp\left(-i  \frac{\theta}{2}\sigma^x\right),\indent R_{\mathrm{Z}}(\theta)=\exp\left(-i  \frac{\theta}{2}\sigma^z\right).
\end{equation}
By switching on the coupling between neighbouring qubit $i$ and $j$, one can implement the $\mathrm{iSWAP}$ gate between qubit $i$ and $j$~\cite{schuch2003natural} with
\begin{equation}
	\mathrm{iSWAP}=\left(
	\begin{matrix}
		1&0&0&0\\
		 0&0&i&0\\
		 0&i&0&0\\
		 0&0&0&1
	\end{matrix}
	\right)=
	\exp\left(-i H_{ij}\frac{3\pi}{2g_{ij}}\right),
\end{equation}
\begin{equation}
	H_{ij}=\frac{\omega}{2}\sigma_i^z+\frac{\omega}{2}\sigma_j^z+g_{ij}(\sigma_i^x\sigma_j^x+\sigma_i^y\sigma_j^y).
\end{equation}
Thus, the single-qubit rotation gates $R_{\mathrm{X}}(\theta)$ and $R_{\mathrm{Z}}(\theta)$ and two-qubit $\mathrm{iSWAP}$ gate can be realized naturally by the physical system with nearest-neighbor switchable XY interactions.

We now review the quantum gates included in the quantum circuit for preparing the DFS basis states as shown in Fig.~\ref{fig1}. It is noted that only a fixed set of quantum gates are included for any $N$ qubits. The quantum gates introduced in the circuit are
\begin{equation}
	\label{eq17}
	\{\mathrm{X},\mathrm{Z},\mathrm{H},\mathrm{CNOT},\mathrm{CZ},\mathrm{CH},\mathrm{CCX},\mathrm{S_1}\},
\end{equation}
where $\mathrm{X}$ is the pauli-X gate, $\mathrm{Z}$ is the pauli-Z gate, $\mathrm{H}$ is the Hadamard gate, $\mathrm{CNOT}$ is the two-qubit controlled-NOT gate, $\mathrm{CZ}$ is the two-qubit controlled-Z gate, $\mathrm{CH}$ is the two-qubit controlled-Hadamard gate, $\mathrm{CCX}$ is the three-qubit Toffoli gate, and $\mathrm{S_1}=I-2|11\cdots 1\rangle\langle 11\cdots 1|$ is the multi-qubit phase gate. Example quantum circuits for preparing two basis states in the $N=4$ case are shown in Fig.~\ref{fig3} and Fig.~\ref{fig4}, respectively. We now show how these quantum gates can be realized with the physical system as in Eq.~(\ref{eq13}), using only $R_{\mathrm{X}}(\theta)$, $R_{\mathrm{Z}}(\theta)$, and $\mathrm{iSWAP}$ gates.
\begin{figure*}[t]
	\includegraphics[width=\linewidth]{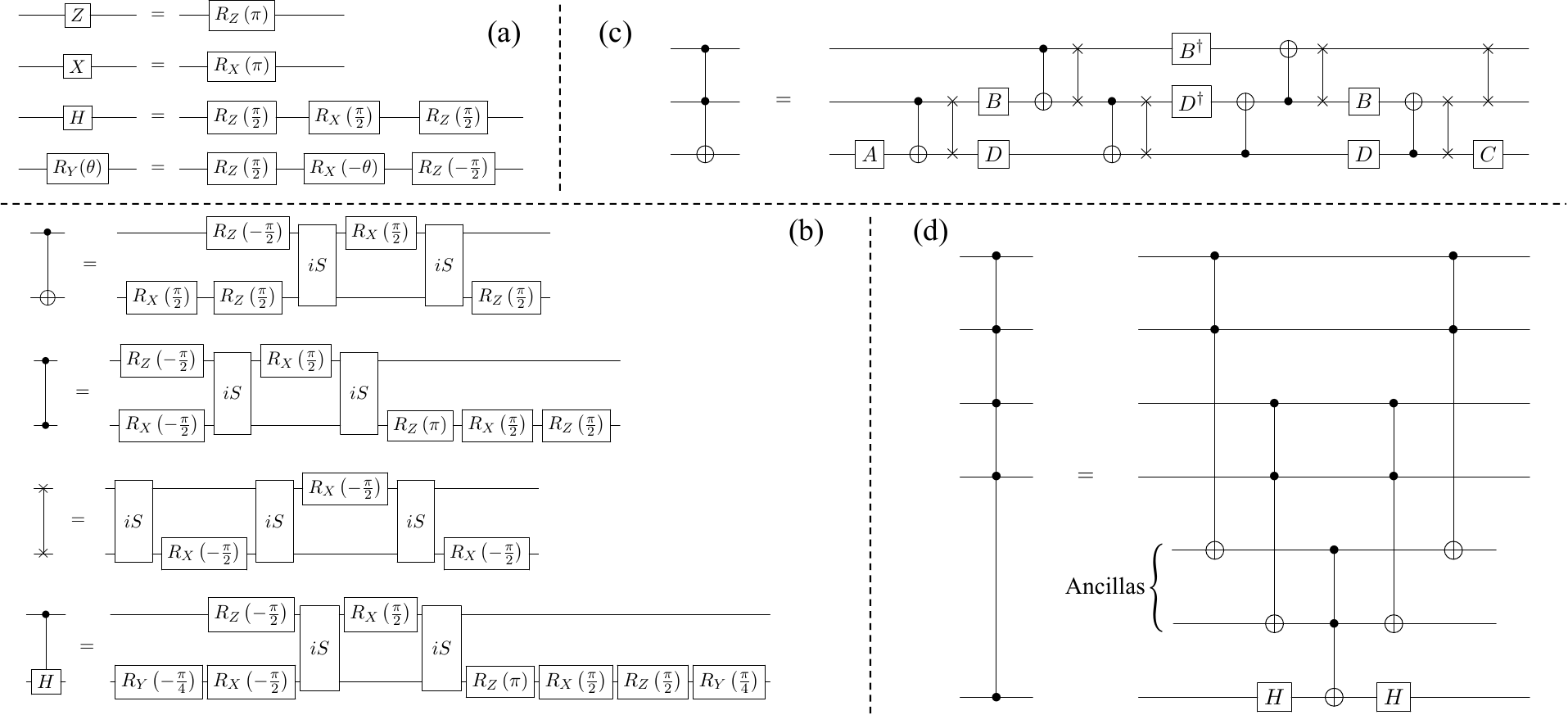}
	\caption{The quantum circuit for realizing the quantum gates in Eq.~(\ref{eq17}). (a) Implementation of single-qubit gates with $R_{\mathrm{X}}(\theta)$ and $R_{\mathrm{Z}}(\theta)$ gates. (b) Implementation of two-qubit gates with $R_{\mathrm{X}}(\theta)$, $R_{\mathrm{Z}}(\theta)$ and $\mathrm{iSWAP}$ gates. $iS$ in the figure denotes the $\mathrm{iSWAP}$ gate. (c) Implementation of Toffoli gate with single-qubit gates and two-qubit gates. The value of single-qubit gates $A$, $B$, $C$, $D$ in (c) is given in Appendix.~\ref{ap7}. (d) Implementation of multi-qubit phase gate with a series of Toffoli gates by introducing ancillary qubits.}
	\label{fig7}
\end{figure*}

First, we discuss how to realize single-qubit gates. Single-qubit gates $\mathrm{X}$ and $\mathrm{Z}$ are equivalent to rotation gates up to a global phase as
\begin{equation}
	\mathrm{X}=\exp\left(i\frac{\pi}{2}\right)R_{\mathrm{X}}(\pi),\indent \mathrm{Z}=\exp\left(-i\frac{\pi}{2}\right)R_{\mathrm{Z}}(\pi),
\end{equation}
and $\mathrm{H}$ can be decomposed with rotation gates following Ref.~\cite{nielsen2010quantum} as
\begin{equation}
	\label{eq15}
	\mathrm{H}=\exp\left(i\frac{\pi}{2}\right)R_{\mathrm{Z}}\left(\frac{\pi}{2}\right)R_{\mathrm{X}}\left(\frac{\pi}{2}\right)R_{\mathrm{Z}}\left(\frac{\pi}{2}\right).
\end{equation}
Thus, all single qubit gates used in the quantum circuit for preparing DFS basis states can be realized as shown in Fig.~\ref{fig7}(a).

Next, we discuss how to realize two-qubit gates and $\mathrm{CCX}$ gate. The decomposition of $\mathrm{CNOT}$ gate is discussed in Ref.~\cite{schuch2003natural}. For the $\mathrm{CZ}$ gate, we have
\begin{equation}
	\mathrm{CZ}=(I\otimes \mathrm{H})\mathrm{CNOT} (I\otimes \mathrm{H}),
\end{equation}
thus the $\mathrm{CZ}$ gate can be realized via the decomposition of Hadamard gates and $\mathrm{CNOT}$ gates. For the $\mathrm{CH}$ gate, we can decompose the $\mathrm{CH}$ gate as
\begin{align}
	\label{eq16}
	\mathrm{CH}=\left(I\otimes R_{\mathrm{Y}}\left(\frac{\pi}{4}\right)\right)\mathrm{CZ}\left(I\otimes R_{\mathrm{Y}}\left(-\frac{\pi}{4}\right)\right).
\end{align}
And the single-qubit rotation-Y gate $R_{\mathrm{Y}}(\theta)$ can be represented by $R_{\mathrm{X}}(\theta)$ and $R_{\mathrm{Z}}(\theta)$ as
\begin{equation}
	\label{eq19}
	R_{\mathrm{Y}}(\theta)=R_{\mathrm{Z}}\left(-\frac{\pi}{2}\right)R_{\mathrm{X}}\left(-\theta\right)R_{\mathrm{Z}}\left(\frac{\pi}{2}\right).
\end{equation}
Considering $\mathrm{CZ}$ can be decomposed with single-qubit gates and $\mathrm{iSWAP}$ gate, thus following Eq.~(\ref{eq16}) and Eq.~(\ref{eq19}), we can represent the $\mathrm{CH}$ gate with with single-qubit gates and $\mathrm{iSWAP}$ gate. Thus, all two-qubit gates included in our quantum circuit can be realized with only $R_{\mathrm{X}}(\theta)$, $R_{\mathrm{Z}}(\theta)$, and $\mathrm{iSWAP}$ gates, which is shown in Fig.~\ref{fig7}(b). Following the decomposition of the Toffoli gate as in Ref.~\cite{schuch2003natural}, we can decompose the Toffoli gate with single qubit gates, $\mathrm{CNOT}$ gate, and $\mathrm{SWAP}$ gate as shown in Fig.~\ref{fig7}(c). Thus, the Toffoli gate can be realized based on the realization of single-qubit gates and two-qubit gates, all of which can be realized with only $R_{\mathrm{X}}(\theta)$, $R_{\mathrm{Z}}(\theta)$, and $\mathrm{iSWAP}$ gates.

Finally, we discuss how to implement the multi-qubit phase gate $\mathrm{S_1}=I-2|11\cdots 1\rangle\langle 11\cdots 1|$. Following Ref.~\cite{nielsen2010quantum}, multi-qubit phase gate can be realized with the Toffoli gates by introducing ancillary qubits. Generally, to implement an $n$-qubit phase gate, a total number of $n-3$ ancillary qubits and $2n-5$ Toffoli gates are needed. An implementation of a $5$-qubit phase gate with the Toffoli gates is shown in Fig.~\ref{fig7}(d). As the Toffoli gates can be realized with only $R_{\mathrm{X}}(\theta)$, $R_{\mathrm{Z}}(\theta)$, and $\mathrm{iSWAP}$ gates as discussed above, the multi-qubit phase gate can also be realized. It is noted that this is a general implementation method for multi-qubit phase gate, for certain number of qubits or physical platform, there also exist some experimental realizations that require fewer gates with simpler implementation~\cite{zhang2024realization,alex2021realization,cai2021all,deng2007implemtation,lin2006one,xu2022one}. It is also noted that the fundamental two-qubit gate used here  is the $\mathrm{iSWAP}$ gate, there are also some other methods for realizing a set of universal gates using general $XY$ entangling gates~\cite{abrams2020implementation}, or $\mathrm{CNOT}$ gates~\cite{vidal2004universal,shende2004minimal} as fundamental two-qubit gates, which may suit better for different physical systems.

We summarize the resource cost for implementing the quantum gates in Eq.~(\ref{eq17}) in Table.~\ref{table2} by showing the number of two-qubit $\mathrm{iSWAP}$ gates and the time needed, which is calculated considering the gate operations are run parallely. Specifcally, the frequency $\omega$ of transmon qubit satisfies $\omega/2\pi\in[4\,\mathrm{GHz}, 8\,\mathrm{GHz}]$, the Rabi frequency $\Omega$ of the pulse for implementing the single-qubit rotation gates satisfies $\Omega/2\pi\in[10\,\mathrm{MHz},100\,\mathrm{MHz}]$, and the coupling strength $g$ satisfies $g/2\pi\in[5\,\mathrm{MHz}, 100\,\mathrm{MHz}]$~\cite{koch2007charge}. To estimate the time for implementing these quantum gates, we take the frequency of the transmon qubit to be $\omega/2\pi=6\,\mathrm{GHz}$, the Rabi to be $\Omega/2\pi=25\,\mathrm{MHz}$, and the coupling strength to be $g/2\pi=25\,\mathrm{MHz}$.
\begin{table}[b]
	\centering
	\caption{Resource cost for implementing quantum gates.}
	\begin{ruledtabular}
		\begin{tabular}{c|cc}
			Gate & Needed $\mathrm{iSWAP}$ number & Needed time\\
			\hline
			$\mathrm{Z}$ & 0 & neglectable\\
			$\mathrm{X}$ & 0& $10\,$ns\\
			$\mathrm{H}$ & 0 & $5\,$ns\\
			$\mathrm{CNOT}$ & 2 & $70\,$ns\\
			$\mathrm{CZ}$ & 2 & $85\,$ns\\
			$\mathrm{CH}$ & 2 & $105\,$ns\\
			$\mathrm{CCX}$ & 10 & $400 \sim 500\,$ns\\
			$5$-qubit $\mathrm{S_1}$ & 50 & $1.2 \sim 1.5\,\mu$s\\
			$n$-qubit $\mathrm{S_1}$ & $20n-50$ & less than $0.4n \,\mu$s
		\end{tabular}
	\end{ruledtabular}
	\label{table2}
\end{table}

Finally, we map the quantum circuit for preparing $4-$qubit DFS basis states as shown in Fig.~\ref{fig3} and Fig.~\ref{fig4} to a physcial system with $7$ qubits with the structure of coupling as shown in Fig. Each quantum gate in the quantum circuit is decomposed as realizable quantum gates, i.e., the single-qubit rotation gate $R_{\mathrm{X}}(\theta)$, $R_{\mathrm{Z}}(\theta)$ and the two-qubit $\mathrm{iSWAP}$ gate. The estimated running time of the decomposed quantum circuit as in Fig.~\ref{fig3} is $80$ ns, and the estimated running time of the decomposed quantum circuit as in Fig.~\ref{fig4} is $4.5\,\mu$s. The running time required is significantly lower than the coherence time of the transmon qubit, which is over $100\,\mu$s~\cite{koch2007charge,gu2017microwave}. Thus, the resource cost by our approach can be satisfied with NISQ superconducting hardware capabilities and our approach can be realized on NISQ superconducting devices.

\begin{figure}[t]
	\includegraphics[width=\linewidth]{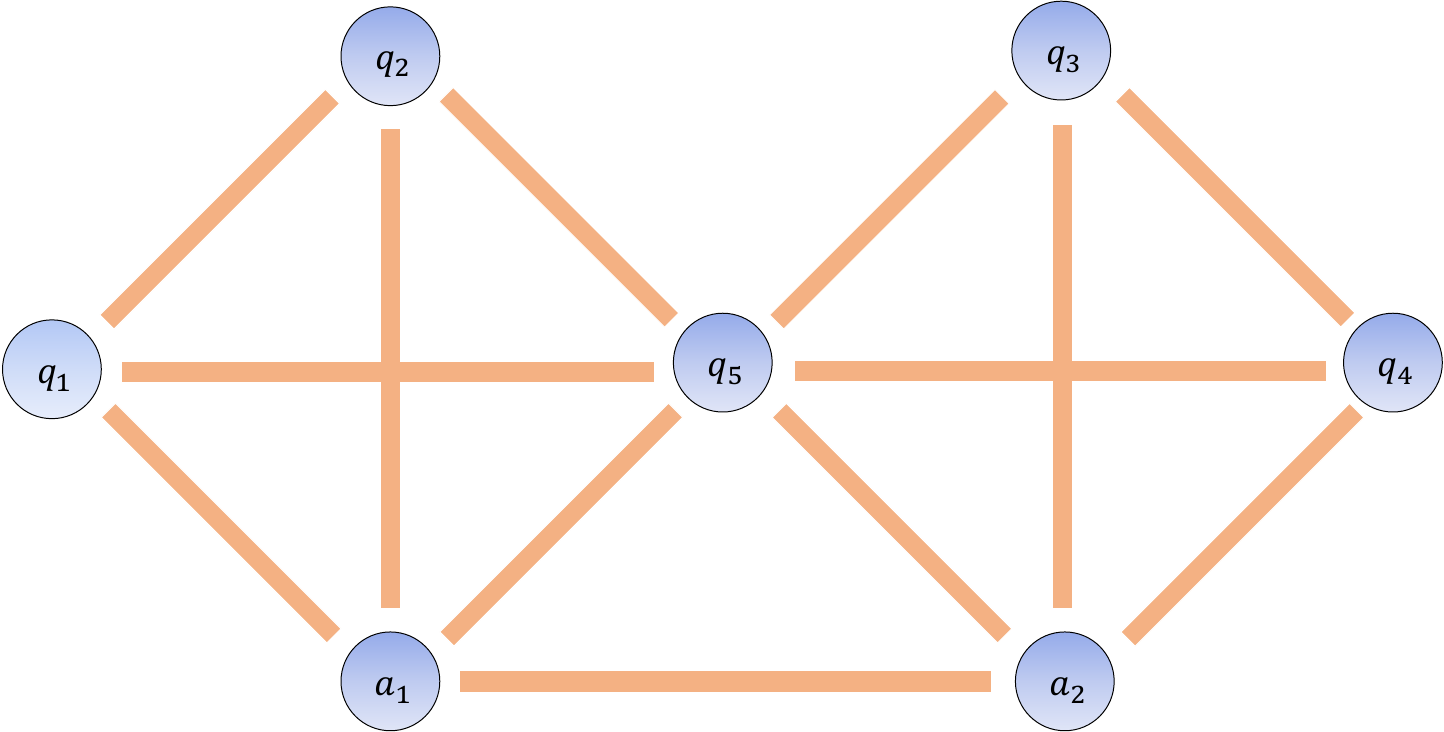}
	\caption{The structure of coupling to prepare the $4$-qubit DFS basis states following the quantum circuit in Fig.~\ref{fig4}. The blue dots represent the transmon qubits, and the orangle lines represent a switchable coupling between the qubits. The definition of qubits $q_0,q_1,q_2,q_3,q_4$ is the same as Fig.~\ref{fig4}, and qubits $a_0,a_1$ are the ancillary qubits for implementing the $5$-qubit phase gates.}
	\label{fig8}
\end{figure}

\section{Discussions and Conclusions}
\label{sec5}
In this work, we propose a general method to prepare all basis states of decoherence-free subspaces (DFSs). This approach is universally applicable to DFSs of arbitrary dimensions and across diverse physical implement platforms of quantum information processing. By iteratively applying the quantum circuit described in Fig.~\ref{fig1}, we can prepare a complete set of orthogonal basis states spanning the DFS.

The computational complexity of our method scales as $O(d(N)^2 \kappa^2\ln(\kappa/\epsilon))$ and the number of qubits needed is $O(d(N))$, where $d(N)$ denotes the dimension of the DFS for $N$ qubit systems, $\kappa$ represents the condition number of $A_{d(N)}$ as defined in Section.~\ref{sec2}, and $\epsilon$ is the infidelity of the prepared bases. This complexity for preparing $d(N)$ quantum states only increases with the factor of $d(N)^2$ and we do not need to calculate the classical vector form of each basis in advance. And the complexity is only logarithmic to $\epsilon^{-1}$ which is the the infidelity of state preparation. Thus, resource-efficient preparation of high-fidelity orthogonal bases is guaranteed. Another obvious but trivial way for preparing the basis states of the DFS is that we calculate the classical vector form of each basis state in advance, and then use the method of general quantum state preparation methods~\cite{long2001efficient,zhang2022quantum,plesch2011quantum}  to prepare these basis states. Calculating the all $d(N)$ classical matrix form for $d(N)$ basis states with the method in Ref.~\cite{cabello2007six} has the time complexity $O(2^Nd(N))$, and preparing all $d(N)$ basis states with the previously best methods as in~\cite{zhang2022quantum} has the time complexity $\Theta(Nd(N))$, requiring $O(2^N)$ ancillary qubits. Combining these two terms, the total complexity of this method is $O(2^Nd(N))$ with the requirement of $O(2^N)$ qubits. This cost of resource is exponenetially higher than that of our method, thus we claim that our method is efficient for preparing the set of orthogonal complete bases of the DFS.

The qubits and the circuit depth required for preparing DFS basis states both scale $O(d(N))$, which are linear to the number of DFS basis states. Thus, for systems containing small number of qubits, e.g., $N=4,\,6,\,,8,\,10$, the preparation of all the basis states of the DFS can be realized with our methods on current superconducting qubit system composed with transmon qubits and tunable couplers, considering the available coherence time of current quantum hardware capabilities~\cite{google2020hartree,google2023suppressing,acharya2024quantum}. That is, our method can be demostrated with superconducting processors in the NISQ era, and provide a pathway for realizing future fault-tolerant quantum computation.

It is also worth mentioning that all our simulations are run on a noiseless simulator, and the noise is not considered for physcial realization. Thus, when the proposed method is realized experimentally on a physical platform with noise, the fidelity of the prepared final state may be limited by the scale of noise in the real-physical system. Meanwhile, to prepare the basis states of a large DFS with exponentially increased number of basis states, the running time required by our method is beyond the coherence time limit in the NISQ era. Thus, we also look forward to the breakthrough in accessing physical systems with smaller scale of noise and longer coherence time, thus making the proposed method more reliable for larger physcial systems, which are capable of encoding a larger number of logical qubits in the DFS.

It is noted that one of the important assmption of the DFS is that all $N$ qubits share the same environment. Many types of quantum processes can be modeled as this kind of noise, which identically couples with each qubit, and one of these quantum processes is the cosmic-ray-induced quantum noise, which was measured and modelled in superconducting qubit systems~\cite{Li2025}. It was found that cosmic-ray-induced quantum noise is the kind of noise that identically interacted with each qubit, thus highlighting the practice value of DFS as a quantum code for suppressing the quantum noise in a real physical system.

We finally emphasize that the proposed method not only can be used to prepare the DFS basis states, but also can be considered as a general method for orthogonalizing a set of quantum states. The query complexity of the proposed method is lower than currrently best quantum algorithms for states orthogonalization~\cite{li2025quantumalgorithmvectorset}. Thus, the proposed method can be used for solving the problem of Gram-Schmidt orthogonalization, matrix QR decomposition, and many other problems in linear algebra that base on vector orthogonalization~\cite{horn2012matrix, bhatia2013matrix,parlett2000qr,leon2013gram}.

\begin{acknowledgments}
	This work was supported by Innovation Program for Quantum Science and Technology (Grant No. 2021ZD0300201).
\end{acknowledgments}

\appendix
\onecolumngrid
\newpage
\section{Introduction to decoherence free subspace}
\label{ap1}
In this section, we briefly introduce the motivation and the concept of the decoherence free subspace, mainly based on the very first paper of the noiseless quantum code~\cite{zanardi1997noiseless}.

The core of quantum computing lies in using quantum devices for information storage and processing. By encoding classical information onto quantum systems and operating and measuring these quantum systems, we can achieve the storage, processing, and retrieval of information, thereby realizing quantum computing. However, when using quantum devices to store and process classical information, one of the most significant challenges is that quantum systems continuously interact with their surrounding environment, resulting in the leakage and loss of information from the quantum system to the environment, ultimately leading to an increase in the entropy and a decrease in the fidelity of the information processed and retrieved through the quantum device. This process is also known as quantum noise and is one of the biggest obstacles in the development and application of quantum computers~\cite{nielsen2010quantum}.

To solve this problem, the concept of quantum error-correcting codes was proposed. Quantum error-correcting codes are conceptually similar to classical error-correcting codes, using multiple physical qubits to encode a small number of logical qubits, thereby increasing the redundancy of the information and reducing the error rate of the logical qubits, achieving greater resistance to noise. Since Shor proposed the 9-qubit Shor code, designing various types of quantum error-correcting codes has always been one of the research hotspots. Researchers have been dedicated to proposing quantum codes with larger code distance, higher encoding efficiency, and higher error threshold. Currently, surface-code-based quantum error-correcting codes are the state-of-the-art quantum error-correcting codes available. The surface code uses $n$ physical qubits to encode one logical qubit, with a code distance $\sqrt{n}$ and an error threshold of approximately $1\%$.

However, the error detection and correction technology based on quantum error-correcting codes is an active error-correction strategy that requires continuous measurement and correction during the operation of the quantum circuit. Therefore, this strategy is harder to be implemented on the existing NISQ devices with short coherence time and high error rate of physical qubits. Thus, a passive error-correction is proposed which is known as the decoherence free subspace. First, we assume that the physical system is consist of $n$ physical qubits with the same common environment. Under this assumption, in the Hilbert space of dimension $2^n$, there exists a subspace where the quantum state composed of the $n$ physical qubits and the environment are decoupled. Therefore, the quantum state  can be immune to the noise brought by the environment. This subspace is called the decoherence free subspace. Therefore, when we regard the bases of the DFS as the bases of logical qubits and encode quantum information onto them, the quantum information is inherently immune to the noise caused by the environment. Thus, without the need for continuous measurements and error correction, longer coherence time and lower error rate of logical qubits can be achieved. Next, we give a mathematical description of the DFS theory.

We denote the physical system containing $n$ spin-$1/2$ particles and their common environment(bath) as $S$ and $B$, respectively, where we assume the $n$ physical qubits share the same environment. The environment is described with a single  thermal bath described by a collection of noninteracting linear oscillators. Thus, the Hamiltonian of the $n$ spin-$1/2$ particles and the environment is denoted as
\begin{equation}
	H_S=E_S\sum_{i=1}^n\sigma_z^i
\end{equation}
and
\begin{equation}
	H_B=\sum_k \omega_k b_k^{\dagger}b_k,
\end{equation}
respectively, where $E_S$ is the energy of a single particle and $b_k$ is the annihilation operator for the $k$th mode of the environment. The interaction between the environment and the physical system is
\begin{equation}
	H_I=\sum_{k,i=1}^n\left(g_{ki}\sigma_i^{+}b_k+f_{ki}\sigma_i^{+}b_k^{\dagger}+h_{ki}\sigma_z^ib_k+\mathrm{H.c.}\right),
\end{equation}
where the first term describes the excitation of the qubit by the absorption of a bath mode with probability amplitude $g_{ki}$, the second term is the rotating wave term, and the third term describes the dephasing process of the qubit due to its coupling with the environment. Following the assumption that all $n$ physical qubits share the same common environment, $g_{ki},f_{ki}$ and $h_{ki}$ are independent of $i$. Thus, if we denote $S_{+}=\sum_{i=1}^n\sigma_i^+$, $S_{-}=\sum_{i=1}^n\sigma_i^-$, and $S_{z}=\sum_{i=1}^n\sigma_i^z$, we can rewrite the interaction Hamiltonian to be
\begin{equation}
	H_I=\sum_{k}\left(g_{k}S_{+}b_k+f_{k}S_{+}b_k^{\dagger}+h_{k}S_zb_k+\mathrm{H.c.}\right),
\end{equation}
thus the Hamiltonian of the whole system combining the system and the environment is
\begin{align}
	\label{eqa4}
	H&=H_S+H_B+H_I\nonumber\\
	&=E_SS_z+\sum_k \omega_k b_k^{\dagger}b_k\nonumber\\
	&+\sum_{k}\left(g_{k}S_{+}b_k+f_{k}S_{+}b_k^{\dagger}+h_{k}S_zb_k+\mathrm{H.c.}\right).
\end{align}

Generally, suppose the system and the environment have a collective initial state $\rho_S\otimes\rho_B$, following the evolution operator $U(t)=\exp(-iHt)$ for some time $t$ and the Hamiltonian $H$, the final state of the physical system is
\begin{equation}
	\rho_S^{'}=tr_B(U(t)\rho_S\otimes \rho_B U(t)^{\dagger}),
\end{equation}
where $tr_B$ means taking the partial trace over the environment. Usually $\rho_S^{'}\neq \rho_S$, which indicates there is error on the physical system due to its coupling with the environment.

However, let's consider the special case that the system is initialized with
\begin{equation}
	\rho_S=\sum_{i,j}\rho_{ij}|\psi_i\rangle\langle \psi_j|,
\end{equation}
where each $|\psi_i\rangle$ and $|\psi_j\rangle$ lie in the kernel space of the total spin operator $S^2$,
\begin{equation}
	S^2|\psi_i\rangle=0\quad S^2|\psi_j\rangle=0.
\end{equation}
The total spin operator is defined as
\begin{equation}
	S^2=S_x^2+S_y^2+S_z^2=S_+^2+S_-^2+S_z^2,
\end{equation}
where each $S_{\alpha}=\sum_{i=1}^ns_{\alpha}^i$, $\alpha=x,y,z,+,-$. And without loss of generality, the initial state of the environment can be denoted as
\begin{equation}
	\rho_B=\sum_{KK'}b_{KK'}|K\rangle\langle K'|,
\end{equation}
where each $|K\rangle$ and $|K'\rangle$ are eigenvectors of the environment Hamiltonian $H_B$ with eigenvalues $K$ and $K'$, respectively. In this case, let's calculate the final state of the physical system with evolution operator $U(t)=\exp(-iHt)$ for some time $t$ and the Hamiltonian $H$ as defined in Eq.~(\ref{eqa4}).

The initial state is
\begin{equation}
	\rho_S\otimes \rho_B=\sum_{ijKK'}\rho_{ij}b_{KK'}(|\psi_i\rangle|K\rangle)(\langle K'|\langle \psi_j|).
\end{equation}
For each $|\psi_i\rangle|K\rangle$, we have
\begin{align}
	H|\psi_i\rangle|K\rangle&=H_S|\psi_i\rangle|K\rangle+H_B|\psi_i\rangle|K\rangle+H_I|\psi_i\rangle|K\rangle.
\end{align}
Using $S^2|\psi_i\rangle=0$, we know $S_{\alpha}|\psi_i\rangle=0$, $\alpha=x,y,z,+,-$. Thus,
\begin{equation}
	H_S|\psi_i\rangle|K\rangle=E_SS_z|\psi_i\rangle|K\rangle=0,
\end{equation}
and
\begin{align}
	H_I|\psi_i\rangle|K\rangle&=\sum_{k}(g_{k}S_{+}b_k+f_{k}S_{+}b_k^{\dagger}\nonumber\\
	&+h_{k}S_zb_k+\mathrm{H.c.})|\psi_i\rangle|K\rangle\nonumber\\
	&=0.
\end{align}
We have $H|\psi_i\rangle|K\rangle=H_B|\psi_i\rangle|K\rangle=K|\psi_i\rangle|K\rangle$. Thus,
\begin{align}
	U(t)|\psi_i\rangle|K\rangle&=\exp(-iHt)|\psi_i\rangle|K\rangle\nonumber\\
	&=\sum_{l=0}^{\infty}\frac{(-it)^l}{l!}H^l|\psi_i\rangle|K\rangle\nonumber\\
	&=\sum_{l=0}^{\infty}\frac{(-it)^l}{l!}K^l|\psi_i\rangle|K\rangle\nonumber\\
	&=\exp(-iKt)|\psi_i\rangle|K\rangle,
\end{align}
and $\langle K'|\langle \psi_j|U(t)^{\dagger}=\langle K'|\langle \psi_j|\exp(iKt)$. Thus the final state of the system and the environment is
\begin{align}
	&U(t)\rho_S\otimes \rho_BU(t)^{\dagger}\nonumber\\
	=&\sum_{ijKK'}\rho_{ij}b_{KK'}U(t)|\psi_i\rangle|K\rangle\langle K'|\langle \psi_j|U(t)^{\dagger}\nonumber\\
	=&\sum_{ijKK'}\rho_{ij}b_{KK'}\exp(-i(K-K')t)|\psi_i\rangle|K\rangle\langle K'|\langle \psi_j|.
\end{align}
Taking the partial trace over the environment, the final state $\rho_S^{'}$ of the physical system satisfies
\begin{align}
	\rho_S^{'}&=tr_B\left(\sum_{ijKK'}\rho_{ij}b_{KK'}\exp(-i(K-K')t)|\psi_i\rangle|K\rangle\langle K'|\langle \psi_j|\right)\nonumber\\
	&=\sum_{ijKK'}\rho_{ij}b_{KK'}\exp(-i(K-K')t)|\psi_i\rangle\langle \psi_j|\sum_{K_0}\langle K_0|K\rangle\langle K'|K_0\rangle\nonumber\\
	&=\sum_{ijKK'}\rho_{ij}b_{KK'}\exp(-i(K-K')t)|\psi_i\rangle\langle \psi_j|\sum_{K_0}\delta_{K_0K}\delta_{K_0K'}\nonumber\\
	&=\sum_{ij}\rho_{ij}\sum_{K_0}b_{K_0K_0}|\psi_i\rangle\langle \psi_j|\nonumber\\
	&=\sum_{ij}\rho_{ij}|\psi_i\rangle\langle \psi_j|=\rho_S.
\end{align}

Thus, we have proved that for all physical systems in the quantum states of $\rho_S=\sum_{ij}\rho_{ij}|\psi_i\rangle\langle \psi_j|$ with $S^2|\psi_i\rangle=0, S^2|\psi_j\rangle=0$, the systems are decoupled with the environment. After a period of free evolution of the whole system combining the physical system and the environment, the quantum state of the physical system will not be disrupted by the environment, and the quantum information stored in the physical system will not be leaked to the environment, as the final state and the initial state of the physical system are equal. Thus this type of quantum state can be immune to the environmental noise. All the states $|\psi\rangle$ that satisfy $S^2|\psi\rangle=0$ form a subspace in the whole Hilbert space, which is called the decoherence free subspace. For a physical system containing $n$ spin-$1/2$ particles, the dimension of the Hilbert space is $2^n$ and it is proved~\cite{zanardi1997noiseless} that the dimension of the DFS is
\begin{equation}
	d(n)=\frac{n!}{(n/2)!(n/2+1)!}=O(\frac{2^n}{n^{3/2}}).
\end{equation}

\section{Construction of complete bases of DFS}
\label{ap2}
In this section, we provide a method that generate a set of complete bases $\{|a_1\rangle,|a_2\rangle, \cdots,|a_{d(N)}\rangle\}$ for the $N$-qubit decoherence-free subspace. Following the main text, each $|a_i\rangle$ has a form of
\begin{equation}
	\label{eqa1}
	|a_i\rangle=|\psi_{i_1i_2}\rangle\otimes|\psi_{i_3i_4}\rangle\otimes\cdots\otimes |\psi_{i_{N-1}i_{N}}\rangle,
\end{equation}
where $\{i_1,i_2,\cdots,i_{N}\}$ is a permutation of $\{1,2,\cdots,N\}$, $\forall i=1,2,\cdots, d(N)$ and $|\psi_{i_mi_{m+1}}\rangle$ denotes a subsystem in the singlet state containing the $i_m$th qubit and the $i_{m+1}$th qubit. We show how to permute each $\{i_1,i_2,\cdots,i_{N}\}$ so that $\{|a_1\rangle,|a_2\rangle, \cdots,|a_{d(N)}\rangle\}$ is a set of complete bases for the $N$-qubit DFS.

It is noted that the dimension $d(N)$ for the $N$-qubit DFS is
\begin{equation}
	d(N)=\frac{N!}{(N/2)!(N/2+1)!},
\end{equation}
also known as the Catalan number, which occurs in many counting problems. The Catalan number has a recurrence form:
\begin{equation}
	d(N)=\sum_{i=0}^{N-1}d(i)d(N-1-i),
\end{equation}
which is often more commonly used. One of these counting problems whose solution involves the Catalan number is to count the numbers of valid parentheses sequences, from which we can construct the set of $\{|a_1\rangle,|a_2\rangle, \cdots,|a_{d(N)}\rangle\}$.

We first introduce the problem of counting the numbers of valid parentheses sequences. The enumeration and generation of valid parentheses sequences constitute a fundamental problem in combinatorics and computer science, with profound implications for algorithmic design and formal language theory. A sequence of parentheses is deemed valid if it satisfies two conditions: every opening parenthesis '$($' admits a corresponding closing parenthesis '$)$', and no prefix of the sequence contains more closing parentheses than opening ones. This definition ensures balanced nesting, mirroring syntactic structures in programming languages and algebraic expressions. The number of valid sequences of length $N$, i.e., $N/2$ pairs, is given by the $N$-th Catalan number $d(N)$. For example, for three pairs of parentheses, the number of valid sequences is $d(3)=5$, and the valid sequences are: \texttt{()()()}, \texttt{()(())}, \texttt{(())()}, \texttt{(()())}, \texttt{((()))}.

The reason that the number of valid parentheses sequences follow the Catalan number is mainly because of the recursive construction process of the sequences. The combinatorial structure of valid parentheses sequences admits a unique recursive decomposition. Any non-empty valid sequence $S$ of $N/2$ pairs can be expressed in the canonical form:
\begin{equation}
	\label{eqa2}
	S = (\,A\,)\,B
\end{equation}
where: $A$ is a valid sequence of $i$ pairs with $0 \leq i \leq N/2-1$, $B$ is a valid sequence of $N/2-1-i$ pairs. Thus, if we denote the number of valid parentheses sequences for $N/2$ pairs of parentheses to be $C_N$, then this decomposition establishes the recurrence relation
\begin{equation}
	\label{eqa3}
	C_N = \sum_{i=0}^{N-1} C_i C_{N-1-i} \quad \text{with} \quad C_0 = 1,
\end{equation}
which is the recurrence relation of the Catalan number. Thus we have
\begin{equation}
	C_N=d(N)=\frac{N!}{(N/2)!(N/2+1)!}.
\end{equation}
\begin{table}[!t]
	\centering
	\begin{tabular}{m{2cm}<{\centering}|m{4cm}<{\centering}|m{4cm}<{\centering}}
		\hline
		State Index & Physical state & Parentheses sequence \\
		\hline
		$|a_1\rangle$ & $|\psi_{12}\rangle\otimes|\psi_{34}\rangle\otimes|\psi_{56}\rangle$ & \texttt{()(())} \\
		$|a_2\rangle$ & $|\psi_{12}\rangle\otimes|\psi_{35}\rangle\otimes|\psi_{46}\rangle$ & \texttt{()(())} \\
		$|a_3\rangle$ & $|\psi_{13}\rangle\otimes|\psi_{24}\rangle\otimes|\psi_{56}\rangle$ & \texttt{(())()} \\
		$|a_4\rangle$ & $|\psi_{13}\rangle\otimes|\psi_{25}\rangle\otimes|\psi_{46}\rangle$ & \texttt{(()())} \\
		$|a_5\rangle$ & $|\psi_{14}\rangle\otimes|\psi_{25}\rangle\otimes|\psi_{36}\rangle$ & \texttt{((()))} \\
		\hline
	\end{tabular}
	\caption{An example for constructing the complete bases for 6-qubit DFS with valid parentheses sequences.}
	\label{table1}
\end{table}

There is a one-to-one relationship between the parentheses sequences with $N/2$ pairs of parentheses and the quantum states containing $N/2$-pairs of entangled qubits as in Eq.~(\ref{eqa1}). For each parentheses sequence $S$, we can view it as a string, and each opening parenthesis $'('$ or closing parenthesis $')'$ has an index. We denote the indexes of a string with length $N$ to be $1,2,\cdots,N$ from the leftmost char to the rightmost char. We record all $N/2$ indexes of the opening parentheses, and denote the indexes of the opening parentheses to be the values of $\{i_1,i_3,\cdots,i_{N-1}\}$ as in Eq.~(\ref{eqa1}). And we record all $N/2$ indexes of the closing parentheses, and denote the indexes of the closing parentheses to be the values of $\{i_2,i_4,\cdots,i_{N}\}$ as in Eq.~(\ref{eqa1}). Thus, we have built a one-to-one relationship between a parentheses sequence and a quantum state as in Eq.~(\ref{eqa1}). For example, \texttt{()(())} is a parentheses sequence containing three pairs of parentheses, it corresponds to the quantum state $|\psi_{12}\rangle\otimes|\psi_{35}\rangle\otimes|\psi_{46}\rangle$.

Following this mapping from parentheses sequences to quantum states, we can build a set of complete states for the $N$-qubit DFS. We first construct all $d(N)$ valid parentheses sequences recursively following the recurrence relation of the parentheses sequences as in Eq.~(\ref{eqa2}) and Eq.~(\ref{eqa3}), then map the $d(N)$ parentheses sequences to $d(N)$ quantum states. Finally, we can denote the $d(N)$ quantum states to be $\{|a_1\rangle,|a_2\rangle,\cdots,|a_{d(N)}\rangle\}$ as the set of the complete bases for the $N$-qubit DFS, thus accomplishing the task of complete bases construction. The example for constructing the complete bases for 6-qubit DFS is given in Table.~\ref{table1}.

The overall complexity of the construction procedure combines the complexity of recursively construction of the valid parentheses sequences and the complexity of the mapping procedure. The overall construction procedure for constructing a set of complete bases for $N$-qubit DFS is
\begin{equation}
	O(Nd(N))=O\left(\frac{2^N}{\sqrt{N}}\right)
\end{equation}
\section{Proof of Eq.~(\ref{eq5})}
\label{ap3}
We prove the equation of
\begin{equation}
	\lim_{m\rightarrow \infty}(Q_k)^m=I-\mathcal{P}_k, \forall k=1,2,\cdots, d(N).
\end{equation}
where $Q_k=P_kP_{k-1}\cdots P_1$, $\mathcal{P}_k$ is the projector onto the subspace spanned by $\{|a_1\rangle, \cdots,|a_k\rangle\}$ and each $P_k$ is defined as $P_k=I-|a_k\rangle\langle a_k|$.

Suppose there is a set of orthogonal complete bases for the subspace spanned by $\{|a_1\rangle, \cdots,|a_k\rangle\}$, and we denote the set of the orthogonal complete bases as $\{|t_1\rangle,|t_2\rangle, \cdots,|t_k\rangle\}$. In this case,
\begin{equation}
	\mathrm{span}\{|t_1\rangle,|t_2\rangle, \cdots,|t_k\rangle\}=\mathrm{span}\{|a_1\rangle,|a_2\rangle, \cdots,|a_k\rangle\},
\end{equation}
and
\begin{equation}
	|\langle t_{k_1}|t_{k_2}\rangle|=\delta_{k_1k_2},\forall k_1,k_2=1,\cdots,k.
\end{equation}
And we can expand the set of the orthogonal complete bases of the subspace to a set of orthogonal complete bases of the whole Hilbert space with dimension $2^N$. The expanded set of orthogonal complete bases is denoted as $\{|t_1\rangle,|t_2\rangle, \cdots,|t_k\rangle\,|t_{k+1}\rangle, \cdots,|t_{2^N}\rangle\}$, and
\begin{equation}
	|\langle t_{n_1}|t_{n_2}\rangle|=\delta_{n_1n_2},\forall n_1,n_2=1,2,\cdots,2^N.
\end{equation}

For $|t_j\rangle$ with $j>k$,
\begin{equation}
	|t_j\rangle\notin \mathrm{span}\{|a_1\rangle,|a_2\rangle, \cdots,|a_k\rangle\},
\end{equation}
thus
\begin{equation}
	(I-\mathcal{P}_k)|t_j\rangle=|t_j\rangle-\mathcal{P}_k|t_j\rangle=|t_j\rangle,
\end{equation}
and
\begin{equation}
	Q_k|t_j\rangle=\prod_{i=k}^{1}(I-|a_i\rangle\langle a_i|)|t_j\rangle=|t_j\rangle.
\end{equation}
Therefore, $\lim_{m\rightarrow \infty}Q_k^m|t_j\rangle=|t_j\rangle$, and $(I-\mathcal{P}_k)|t_j\rangle=|t_j\rangle$, $\forall j=k+1,\cdots,2^N$. We conclude that both the operator of $\lim_{m\rightarrow \infty}(Q_k)^m$ and $(I-\mathcal{P}_k)$ have a set of eigenvectors $\{|t_{k+1}\rangle,\cdots,|t_{2^N}\rangle\}$, in which all eigenvectors correspond to the eigenvalue 1.

For $|t_j\rangle$ with $j\leq k$,
\begin{equation}
	|t_j\rangle\in \mathrm{span}\{|a_1\rangle,|a_2\rangle, \cdots,|a_k\rangle\},
\end{equation}
thus
\begin{equation}
	(I-\mathcal{P}_k)|t_j\rangle=|t_j\rangle-\mathcal{P}_k|t_j\rangle=0,
\end{equation}
and
\begin{equation}
	\parallel Q_k|t_j\rangle\parallel =\parallel \prod_{i=k}^{1}(I-|a_i\rangle\langle a_i|)|t_j\rangle\parallel.
\end{equation}
For operator $P_i=I-|a_i\rangle\langle a_i|$ and an arbitrary state $|\psi\rangle$, if $|\langle a_i|\psi\rangle|>0$, then $\parallel P_i|\psi\rangle\parallel <1$, else $\parallel P_i|\psi\rangle\parallel =1$. Thus for operator $Q_k=P_{k}P_{k-1}\cdots P_1$, if $\exists i\in \{1,\cdots,k\}$, $|\langle a_i|\psi\rangle|>0$, then $\parallel Q_k|\psi\rangle\parallel<1$. For the case that we take $|\psi\rangle=|t_j\rangle$ with $\forall j=1,2,\cdots,k$, we know that
\begin{equation}
	|t_j\rangle\in \mathrm{span}\{|a_1\rangle,|a_2\rangle, \cdots,|a_k\rangle\},
\end{equation}
thus $\exists i\in \{1,\cdots,k\}$, $|\langle a_i|t_j\rangle|>0$. We can conclude that
\begin{equation}
	\parallel Q_k|t_j\rangle\parallel =\parallel \prod_{i=k}^{1}(I-|a_i\rangle\langle a_i|)|t_j\rangle\parallel<1.
\end{equation}
Denote the maximum eigenvalue of $Q_k$ whose magnitude is less than 1 to be $\lambda_k$, then
\begin{equation}
	\parallel Q_k|t_j\rangle\parallel \leq |\lambda_k|t_j\rangle|=|\lambda_k|,
\end{equation}
and
\begin{equation}
	\parallel \lim_{m\rightarrow \infty}(Q_k)^m|t_j\rangle \parallel\leq |\lim_{m\rightarrow \infty}\lambda_k^m|t_j\rangle|=0,
\end{equation}
because $|\lambda_k|<1$ and $|\lim_{m\rightarrow \infty}\lambda_k|^m=0$.
Thus, $\forall j=1,\cdots, k$, we have $(I-\mathcal{P}_k)|t_j\rangle=0$ and $\lim_{m\rightarrow \infty}(Q_k)^m|t_j\rangle=0$. We conclude that both the operator of $\lim_{m\rightarrow \infty}(Q_k)^m$ and $(I-\mathcal{P}_k)$ have a set of eigenvectors $\{|t_{1}\rangle,\cdots,|t_{k}\rangle\}$, in which all eigenvectors correspond to the eigenvalue 0.

Summarily, we conclude that both the operator of $\lim_{m\rightarrow \infty}(Q_k)^m$ and $(I-\mathcal{P}_k)$ have a set of eigenvectors $\{|t_{k+1}\rangle,\cdots,|t_{2^N}\rangle\}$, in which all eigenvectors correspond to the eigenvalue 1, and a set of eigenvectors $\{|t_{1}\rangle,\cdots,|t_{k}\rangle\}$, in which all eigenvectors correspond to the eigenvalue 0. As $\{|t_1\rangle,|t_2\rangle,\cdots,|t_{2^N}\rangle\}$ is a set of orthogonal complete bases for the Hilbert space, we can write the operator $\lim_{m\rightarrow \infty}(Q_k)^m$ and $(I-\mathcal{P}_k)$ in a form of spectral decomposition,
\begin{equation}
	I-\mathcal{P}_k=\sum_{j=1}^{k}0*|t_j\rangle\langle t_j|+\sum_{j=k+1}^{2^N}1*|t_j\rangle\langle t_j|=\sum_{j=k+1}^{2^N}|t_j\rangle\langle t_j|,
\end{equation}
and
\begin{equation}
	\lim_{m\rightarrow \infty}(Q_k)^m=\sum_{j=1}^{k}0*|t_j\rangle\langle t_j|+\sum_{j=k+1}^{2^N}1*|t_j\rangle\langle t_j|=\sum_{j=k+1}^{2^N}|t_j\rangle\langle t_j|.
\end{equation}
Thus we have proved
\begin{equation}
	\lim_{m\rightarrow \infty}(Q_k)^m=I-\mathcal{P}_k.
\end{equation}
\section{Proof of the main result}
\label{ap4}
The main result is described in the main text and reviewed as follows. We define a matrix $A_{d(N)}=\left(|a_1\rangle,|a_2\rangle,\cdots,|a_{d(N)}\rangle\right)$, and $\kappa$ as the conditional number of the matrix $A_{d(N)}$, i.e., the ratio of the maximum singular value $\sigma_{\max}$ and the minimum singular value $\sigma_{\min}$. Then we can construct a set of bases $\{|u_1\rangle, |u_2\rangle,\cdots,|u_{d(N)}\rangle\}$ for the DFS consist of $N$ spin-$1/2$ particles with the methods that we propose, using a total number of $O(d(N))$ qubits and
\begin{equation}
	O\left(d(N)^2\kappa^2\ln\left(\frac{\kappa}{\epsilon}\right)\right)
\end{equation}
quantum gates, such that $\forall k=0, 1,\cdots,d(N)-1$ and $\epsilon>0$, it is satisfied that
\begin{equation}
	1-\parallel \langle u_{k+1}| t_{k+1}\rangle \parallel^2<\epsilon.
\end{equation}
Meanwhile, the value of each $m_{k+1}$ is
\begin{equation}
	m_{k+1}=O\left(\ln\left(\frac{\kappa}{\epsilon}\right)\right)
\end{equation}
so that $|a_{k+1}^{(m_{k+1})}\rangle$ can be denoted as $|u_{k+1}\rangle$. Each $|t_{k+1}\rangle$ is defined as
\begin{equation}
	|t_{k+1}\rangle =\frac{|a_{k+1}\rangle-\mathcal{P}_k|a_{k+1}\rangle}{\parallel |a_{k+1}\rangle-\mathcal{P}_k|a_{k+1}\rangle\parallel},
\end{equation}
and $\mathcal{P}_k$ is the projector onto the subspace $span\{|a_1\rangle,|a_2\rangle,\cdots,|a_k\rangle\}$, $ \forall k=0,1,\cdots,d(N)-1$.

Now we prove this result. We calculate the number of quantum gates needed for constructing arbitrary $|u_{k+1}\rangle$, $\forall k=0,1,\cdots,d(N)-1$. We know for a given $\epsilon$, the quantum circuit as in Fig.~\ref{fig1} needs to be run for multiple times to construct a state $|u_{k+1}\rangle$ that is close enough to the state $|t_{k+1}\rangle$. To construct $|u_{k+1}\rangle$, we first need to take the input of the quantum circuit in Fig.~\ref{fig1} as $|0\rangle^{\otimes k}|a_{k+1}^{(0)}\rangle$, where we denote $|a_{k+1}\rangle$ as $|a_{k+1}^{(0)}\rangle$. Then the quantum circuit is run and measured, if the outcomes are all zeros, then a state $|a_{k+1}^{(1)}\rangle$ is constructed, which is then taken as the input of the same quantum circuit. Following this procedure, we can construct a series of states $\left\{|a_{k+1}^{(1)}\rangle,|a_{k+1}^{(2)}\rangle,\cdots,|a_{k+1}^{(m_{k+1})}\rangle\right\}$ till
\begin{equation}
	\label{eqa5}
	1-\parallel \langle a_{k+1}^{(m_{k+1})}| t_{k+1}\rangle \parallel^2<\epsilon.
\end{equation}
is satisfied, then we can denote the state $|a_{k+1}^{(m_{k+1})}\rangle$ to be $|u_{k+1}\rangle$.

We first calculate how large does the value of $m_{k+1}$ needs to be for a given $\epsilon$, and how many times does the quantum circuit as in Fig.~\ref{fig1} needs to be run.

Following the Eq.~(\ref{eq11}), we know
\begin{equation}
	|a_{k+1}^{(m_{k+1})}\rangle=\frac{(Q_k)^{m_{k+1}}|a_{k+1}\rangle}{\parallel (Q_k)^{m_{k+1}}|a_{k+1}\rangle\parallel}
\end{equation}
Denote $|a_{k+1}\rangle$ as a linear combination of
\begin{equation}
	|a_{k+1}\rangle=\mathcal{P}_k|a_{k+1}\rangle+(I-\mathcal{P}_k)|a_{k+1}\rangle,
\end{equation}
then following the definition of $Q_k=P_kP_{k-1}\cdots P_1$, we have
\begin{align}
&\;\;(Q_k)^{m_{k+1}}|a_{k+1}\rangle\nonumber\\
=&\;\;(Q_k)^{m_{k+1}}\mathcal{P}_k|a_{k+1}\rangle+(Q_k)^{m_{k+1}}(I-\mathcal{P}_k)|a_{k+1}\rangle\nonumber\\
=&\;\;(Q_k)^{m_{k+1}}\mathcal{P}_k|a_{k+1}\rangle+(I-\mathcal{P}_k)|a_{k+1}\rangle.
\end{align}
Thus, if we denote the maximum eigenvalue of $Q_k$ that is less than one to be $\lambda_k$, $\lambda_k<1$, then
\begin{align}
	&\;\;\parallel(Q_k)^{m_{k+1}}|a_{k+1}\rangle\parallel\nonumber\\
	=&\;\;\parallel (Q_k)^{m_{k+1}}\mathcal{P}_k|a_{k+1}\rangle+(I-\mathcal{P}_k)|a_{k+1}\rangle\parallel\nonumber\\
	\leq &\;\;\lambda_k^{m_{k+1}}\parallel \mathcal{P}_k|a_{k+1}\rangle\parallel + \parallel(I-\mathcal{P}_k)|a_{k+1}\rangle\parallel\nonumber\\
	\leq &\;\;\lambda_k^{m_{k+1}}+\parallel(I-\mathcal{P}_k)|a_{k+1}\rangle\parallel,
\end{align}
and
\begin{align}
	&\;\;\parallel(Q_k)^{m_{k+1}}|a_{k+1}\rangle\parallel\nonumber\\
	=&\;\;\parallel (Q_k)^{m_{k+1}}\mathcal{P}_k|a_{k+1}\rangle+(I-\mathcal{P}_k)|a_{k+1}\rangle\parallel\nonumber\\
	\geq&\;\; -\lambda_k^{m_{k+1}}\parallel \mathcal{P}_k|a_{k+1}\rangle\parallel + \parallel(I-\mathcal{P}_k)|a_{k+1}\rangle\parallel\nonumber\\
	\geq&\;\; -\lambda_k^{m_{k+1}}+\parallel(I-\mathcal{P}_k)|a_{k+1}\rangle\parallel.
\end{align}
Thus,
\begin{align}
	&\;\;\parallel \langle t_{k+1}|a_{k+1}^{(m_{k+1})}\rangle\parallel\nonumber\\
	=&\;\;\frac{\parallel \langle t_{k+1}|(Q_k)^{m_{k+1}}|a_{k+1}\rangle\parallel}{\parallel (Q_k)^{m_{k+1}}|a_{k+1}\rangle\parallel}\nonumber\\
	=&\;\;\frac{\parallel \langle a_{k+1}|(I-\mathcal{P}_k)(Q_k)^{m_{k+1}}|a_{k+1}\rangle\parallel}{\parallel (Q_k)^{m_{k+1}}|a_{k+1}\rangle\parallel\parallel (I-\mathcal{P}_k)|a_{k+1}\rangle\parallel}\nonumber\\
	= &\;\;\frac{\parallel \langle a_{k+1}|(I-\mathcal{P}_k)(Q_k)^{m_{k+1}}|a_{k+1}\rangle\parallel}{\parallel (Q_k)^{m_{k+1}}|a_{k+1}\rangle\parallel\parallel (I-\mathcal{P}_k)|a_{k+1}\rangle\parallel}\nonumber\\
	=&\;\; \frac{\parallel \langle a_{k+1}|(I-\mathcal{P}_k)(Q_k)^{m_{k+1}}\mathcal{P}_k|a_{k+1}\rangle+\langle a_{k+1}|(I-\mathcal{P}_k)^2|a_{k+1}\rangle\parallel}{\parallel (Q_k)^{m_{k+1}}|a_{k+1}\rangle\parallel\parallel (I-\mathcal{P}_k)|a_{k+1}\rangle\parallel}\nonumber\\
	\geq &\;\;\frac{\parallel \langle a_{k+1}|(I-\mathcal{P}_k)^2|a_{k+1}\rangle\parallel}{\parallel (Q_k)^{m_{k+1}}|a_{k+1}\rangle\parallel\parallel (I-\mathcal{P}_k)|a_{k+1}\rangle\parallel}\nonumber\\
	= & \;\;\frac{\parallel (I-\mathcal{P}_k)|a_{k+1}\rangle\parallel}{\parallel (Q_k)^{m_{k+1}}|a_{k+1}\rangle\parallel}\nonumber\\
	\geq &\;\;\frac{\parallel (I-\mathcal{P}_k)|a_{k+1}\rangle\parallel}{\lambda_k^{m_{k+1}}+\parallel(I-\mathcal{P}_k)|a_{k+1}\rangle\parallel}=1-\frac{\lambda_k^{m_{k+1}}}{\lambda_k^{m_{k+1}}+\parallel(I-\mathcal{P}_k)|a_{k+1}\rangle\parallel}.
\end{align}
Then
\begin{equation}
	\parallel \langle t_{k+1}|a_{k+1}^{(m_{k+1})}\rangle\parallel^2\geq 1-\frac{2\lambda_k^{m_{k+1}}}{\lambda_k^{m_{k+1}}+\parallel(I-\mathcal{P}_k)|a_{k+1}\rangle\parallel}.
\end{equation}
For a given $\epsilon$, we demand $	1-\parallel \langle a_{k+1}^{(m_{k+1})}| t_{k+1}\rangle \parallel^2<\epsilon$, which is
\begin{equation}
	\frac{2\lambda_k^{m_{k+1}}}{\lambda_k^{m_{k+1}}+\parallel(I-\mathcal{P}_k)|a_{k+1}\rangle\parallel}<\epsilon,
\end{equation}
i.e.,
\begin{equation}
	\lambda_k^{m_{k+1}}<\frac{\epsilon \parallel(I-\mathcal{P}_k)|a_{k+1}\rangle\parallel}{2-\epsilon}.
\end{equation}
When we take $m_{k+1}$ to be
\begin{equation}
	m_{k+1}=\lceil \frac{\ln (\epsilon/\kappa)}{\ln \lambda_k}\rceil ,
\end{equation}
where $\kappa$ is the conditional number of matrix $A_{d(N)}$ and $\lceil \cdot\rceil$ is the upper rounding function, we have
\begin{equation}
	\lambda_k^{m_{k+1}}\leq \frac{\epsilon}{\kappa}<\frac{\epsilon \parallel(I-\mathcal{P}_k)|a_{k+1}\rangle\parallel}{2-\epsilon}.
\end{equation}
Thus, when we take $m_{k+1}$ to be $\lceil \frac{\ln (\epsilon/\kappa)}{\ln \lambda_k}\rceil $, Eq.~(\ref{eqa5}) is satisfied. It is noted that we assume
\begin{equation}
	\label{eqa6}
	\kappa\geq\frac{1}{\parallel(I-\mathcal{P}_k)|a_{k+1}\rangle\parallel},
\end{equation}
which we will prove in Appendix.~\ref{ap5}. As $\ln\lambda_k$ is a constant number, we have
\begin{equation}
	m_{k+1}=\lceil \frac{\ln (\epsilon/\kappa)}{\ln \lambda_k}\rceil=O\left(\ln\left(\frac{\kappa}{\epsilon}\right)\right).
\end{equation}

Next we calculate how many quantum gates are needed to construct $|a_{k+1}^{(m_{k+1})}\rangle$, i.e., the state $|u_{k+1}\rangle$. Following the procedure of state construction, we need to construct a series of states $\left\{|a_{k+1}^{(1)}\rangle,|a_{k+1}^{(2)}\rangle,\cdots,|a_{k+1}^{(m_{k+1})}\rangle\right\}$. And when each of these quantum states is constructed, it is required that all the measurements outcomes of ancillary qubits should be zero. Thus, multiple runs and measurements of the quantum circuit are needed to construct each of the states. We first calculate how many times does the quantum circuit as in Fig.~\ref{fig1} need to be run to construct $|a_{k+1}^{(m_{k+1})}\rangle$. Denote $p_{k+1}^{(j)}$ as the probability that all measurement outcomes of ancillary qubits are zero when we construct state $|a_{k+1}^{(j)}\rangle$, and $r_{k+1}^{(j)}$ as the average number that the quantum circuit as in Fig.~\ref{fig1} needs to be run to get one copy of $|a_{k+1}^{(j)}\rangle$, $j=1,2,\cdots,m_{k+1}$, then we can write a recurrence relation between $r_{k+1}^{(j-1)}$ and $r_{k+1}^{(j)}$
\begin{equation}
	r_{k+1}^{(j)}=\frac{1}{p_{k+1}^{(j)}}\left(r_{k+1}^{(j-1)}+1\right),\forall j=1,2,\cdots,m_{k+1},
\end{equation}
where $r_{k+1}^{(0)}=0$ as $|a_{k+1}^{(0)}\rangle=|a_{k+1}\rangle$. Thus, following the recurrence relation, we have
\begin{equation}
	r_{k+1}^{(j)}=\sum_{i_1=1}^{j}\left(\prod_{i_2=i_1}^{j}\frac{1}{p_{k+1}^{(i_2)}}\right).
\end{equation}
Now we calculate $p_{k+1}^{(j)}$. To construct $|a_{k+1}^{(j)}\rangle$, the input state of the quantum circuit as in Fig.~\ref{fig1} is $|0\rangle^{\otimes k}|a_{k+1}^{(j-1)}\rangle$. Following Eq.~(\ref{eq12}), the probability that the outcomes of all ancillary qubits are all zero is
\begin{equation}
	p_{k+1}^{(j)}=\langle a_{k+1}^{(j-1)}|Q_k^TQ_k|a_{k+1}^{(j-1)}\rangle.
\end{equation}
As
\begin{equation}
	|a_{k+1}^{(j-1)}\rangle=\frac{Q_k^{j-1}|a_{k+1}\rangle}{\parallel Q_k^{j-1}|a_{k+1}\rangle\parallel},
\end{equation}
we have
\begin{equation}
	p_{k+1}^{(j)}=\frac{\parallel Q_k^j|a_{k+1}\rangle\parallel^2}{\parallel Q_k^{j-1}|a_{k+1}\rangle\parallel^2}.
\end{equation}
Thus,
\begin{align}
	r_{k+1}^{(j)}&=\sum_{i_1=1}^{j}\left(\prod_{i_2=i_1}^{j}\frac{1}{p_{k+1}^{(i_2)}}\right)\nonumber\\
	&\leq \sum_{i_1=1}^{j}\left(\prod_{i_2=1}^{j}\frac{1}{p_{k+1}^{(i_2)}}\right)\nonumber\\ &=j\prod_{i_2=1}^{j}\frac{\parallel Q_k^{i_2-1}|a_{k+1}\rangle\parallel^2}{\parallel Q_k^{i_2}|a_{k+1}\rangle\parallel^2}=\frac{j}{\parallel Q_k^{j}|a_{k+1}\rangle\parallel^2},
\end{align}
and
\begin{equation}
	r_{k+1}^{(m_{k+1})}\leq \frac{m_{k+1}}{\parallel Q_k^{m_{k+1}}|a_{k+1}\rangle\parallel^2}.
\end{equation}
As we have proved $\parallel(Q_k)^{m_{k+1}}|a_{k+1}\rangle\parallel\geq -\lambda_k^{m_{k+1}}+\parallel(I-\mathcal{P}_k)|a_{k+1}\rangle\parallel$,  $\lambda_k^{m_{k+1}}\leq \epsilon/\kappa$, and $m_{k+1}=\lceil \frac{\ln (\epsilon/\kappa)}{\ln \lambda_k}\rceil$, we have
\begin{align}
	r_{k+1}^{(m_{k+1})}&\leq \frac{m_{k+1}}{\parallel Q_k^{m_{k+1}}|a_{k+1}\rangle\parallel^2}\nonumber\\
	&\leq \frac{m_{k+1}}{-2\parallel(I-\mathcal{P}_k)|a_{k+1}\rangle\parallel\lambda_k^{m_{k+1}}+\parallel(I-\mathcal{P}_k)|a_{k+1}\rangle\parallel^2}\nonumber\\
	&\leq\frac{m_{k+1}}{-2\parallel(I-\mathcal{P}_k)|a_{k+1}\rangle\parallel\epsilon/\kappa+\parallel(I-\mathcal{P}_k)|a_{k+1}\rangle\parallel^2}\nonumber\\
	&\leq \frac{m_{k+1}}{-2\epsilon/\kappa^2+1/\kappa^2}\nonumber\\
	&\leq 2\kappa^2 m_{k+1}.
\end{align}
Again we use $\parallel (I-\mathcal{P}_k)|a_{k+1}\rangle\parallel\geq 1/\kappa$. By pluging the value of $m_{k+1}$ into the equation above, we have
\begin{equation}
		r_{k+1}^{(m_{k+1})}\leq 2\kappa^2 m_{k+1}=2\kappa^2\lceil \frac{\ln (\epsilon/\kappa)}{\ln \lambda_k}\rceil=O\left(\kappa^2\ln\left(\frac{\kappa}{\epsilon}\right)\right).
\end{equation}
Summarily, the total number that the quantum circuit as in Fig.~\ref{fig1} needs to be run for constructing $|u_{k+1}\rangle=|a_{k+1}^{(m_{k+1})}\rangle$ is
\begin{equation}
	r_{k+1}^{(m_{k+1})}=O\left(\kappa^2\ln\left(\frac{\kappa}{\epsilon}\right)\right),
\end{equation}
and the quantum circuit contains a total number of $O(k)$ quantum gates. Thus, the total quantum gates needed for constructing $|u_{k+1}\rangle$ is
\begin{equation}
	O\left(k\kappa^2\ln\left(\frac{\kappa}{\epsilon}\right)\right).
\end{equation}
And to consstruct all $\{|u_1\rangle, |u_2\rangle, \cdots,|u_{d(N)}\rangle\}$, the total number of quantum gates needed is
\begin{equation}
	\sum_{k=0}^{d(N)-1}O\left(k\kappa^2\ln\left(\frac{\kappa}{\epsilon}\right)\right)=O\left(d(N)^2\kappa^2\ln\left(\frac{\kappa}{\epsilon}\right)\right).
\end{equation}
Thus, the main result has been proved.
\section{Proof of Eq.~(\ref{eqa6})}
\label{ap5}
For matrix $A_{d(N)}=\left(|a_1\rangle,|a_2\rangle,\cdots,|a_{d(N)}\rangle\right)$, and $\kappa$ as the conditional number of the matrix $A_{d(N)}$, i.e., the ratio of the maximum singular value $\sigma_{\max}$ and the minimum singular value $\sigma_{\min}$, we have
\begin{equation}
	\label{eqa7}
	\kappa\geq\frac{1}{\parallel(I-\mathcal{P}_k)|a_{k+1}\rangle\parallel}, \forall k=0,1,\cdots, d(N)-1,
\end{equation}
where each $\mathcal{P}_k$ is the projector onto the subspace spanned by $\{|a_1\rangle,|a_2\rangle,\cdots,|a_k\rangle\}$. Now we prove this result.

 As the matrix $A_{d(N)}$ is a full column rank matrix, we can perform the QR decomposition on the matrix $A_{d(N)}$. Denote $A_{d(N)}=QR$ to be the QR decomposition of matrix $A_{d(N)}$. Then we have
 \begin{equation}
 	Q=\left(|t_1\rangle,|t_2\rangle,\cdots,|t_{d(N)}\rangle\right),
 \end{equation}
where each $|t_{k+1}\rangle$ is defined as
\begin{equation}
	|t_{k+1}\rangle =\frac{|a_{k+1}\rangle-\mathcal{P}_k|a_{k+1}\rangle}{\parallel |a_{k+1}\rangle-\mathcal{P}_k|a_{k+1}\rangle\parallel},
\end{equation}
the same as Eq.~(\ref{eq4}). Following the QR decomposition algorithm, the diagonal terms of matrix $R$ is
\begin{align}
	(R)_{k+1,k+1}&=\parallel |a_{k+1}\rangle-\sum_{i=1}^{k}\langle u_i|a_{k+1}\rangle|u_i\rangle \parallel\nonumber\\
	&=\parallel |a_{k+1}\rangle-\mathcal{P}_{k}|a_{k+1}\rangle\parallel, \forall k=0,1,\cdots, d(N)-1.
\end{align}
As the matrix $R$ is an upper triangle matrix, the diagonal terms of $R$ are the eigenvalues of $R$. Denote the maximum eigenvalue of $R$ to be $\lambda_R^{\max}$ and the minimum sigular value to be $\lambda_R^{\min}$, then
\begin{equation}
	\lambda_R^{\max}=1
\end{equation}
\begin{align}
	\lambda_R^{\min}&=\min_{k=0,\cdots,d(N)-1}R_{k+1,k+1}\nonumber\\
	&=\min_{k=0,\cdots,d(N)-1} \parallel |a_{k+1}\rangle-\mathcal{P}_{k}|a_{k+1}\rangle\parallel.
\end{align}
Thus, defining the the conditional number $\kappa_R$ of $R$ to be the ratio of the maximum singluar value $\sigma_R^{\max}$ and the minimum sigular value $\sigma_R^{\min}$, then
\begin{equation}
	\sigma_R^{\max}\geq \lambda_R^{\max},
\end{equation}
\begin{equation}
	\sigma_R^{\min}\geq \lambda_R^{\min}.
\end{equation}
Thus
\begin{align}
	\label{eqa9}
	\kappa_{R}=\frac{\sigma_R^{\max}}{\sigma_R^{\min}}\geq \frac{\lambda_R^{\max}}{\lambda_R^{\min}}=\frac{1}{\min_{k} \parallel |a_{k+1}\rangle-\mathcal{P}_{k}|a_{k+1}\rangle\parallel}.
\end{align}
As $A_{d(N)}$ is a full column-rank matrix with $A_{d(N)}=QR$, and $Q^{\dagger}Q=I$,
\begin{equation}
	A_{d(N)}^{\dagger}A_{d(N)}=R^{\dagger}Q^{\dagger}QR=R^{\dagger}R.
\end{equation}
Thus
\begin{equation}
	\label{eqa11}
	\kappa=\frac{\sigma_{\max}}{\sigma_{\min}}=\frac{\lambda_{\max}(A_{d(N)}^{\dagger}A_{d(N)})}{\lambda_{\min}(A_{d(N)}^{\dagger}A_{d(N)})}=\frac{\lambda_{\max}(R^{\dagger}R)}{\lambda_{\min}(R^{\dagger}R)}=\kappa_{R}.
\end{equation}
Combining Eq.~(\ref{eqa9}), and Eq.~(\ref{eqa11}) together, we have
\begin{align}
	\kappa=\kappa_R&\geq \frac{1}{\min_{k} \parallel |a_{k+1}\rangle-\mathcal{P}_{k}|a_{k+1}\rangle\parallel}\nonumber\\
	&=\frac{1}{\min_{k} \parallel(I-\mathcal{P}_{k})|a_{k+1}\rangle\parallel}.
\end{align}

i.e.,
\begin{equation}
	\kappa\geq\frac{1}{\parallel(I-\mathcal{P}_k)|a_{k+1}\rangle\parallel}, \forall k=0,1,\cdots, d(N)-1.
\end{equation}
Thus we have proved Eq.~(\ref{eqa6}).
\section{Single-qubit gates in Fig.~\ref{fig7}(c)}
\label{ap7}
Following~\cite{schuch2003natural}, we write down the quantum gates $A,B,C,D$. And we calculate thhe decompositions of these quantum gates with single-qubit rotation gates which are shown as follows. The quantum gates $A,B,C,D$ is definede and shown in Fig.~\ref{fig7}(c).
\begin{equation}
	A=\left(
	\begin{matrix}
		1&0\\
		0&i
	\end{matrix}\right)=\exp(i\frac{\pi}{4})R_{\mathrm{Z}}\left(\frac{\pi}{2}\right)
\end{equation}
\begin{equation}
	B=\frac{1}{\sqrt{4-2\sqrt{2}}}\left(
	\begin{matrix}
		1&1-\sqrt{2}\\
		\sqrt{2}-1&1
	\end{matrix}\right)=R_{\mathrm{Z}}\left(\frac{\pi}{2}\right)R_{\mathrm{X}}\left(\frac{\pi}{4}\right)R_{\mathrm{Z}}\left(-\frac{\pi}{2}\right)
\end{equation}
\begin{equation}
	C=\frac{1}{\sqrt{4-2\sqrt{2}}}\left(
	\begin{matrix}
		1&\sqrt{2}-1\\
		i(\sqrt{2}-1)&-i
	\end{matrix}\right)=\exp\left(i\frac{3\pi}{4}\right)R_{\mathrm{Z}}\left(\pi\right)R_{\mathrm{X}}\left(\frac{\pi}{4}\right)R_{\mathrm{Z}}\left(\frac{\pi}{2}\right)
\end{equation}
\begin{equation}
	D=\left(
	\begin{matrix}
		1&0\\
		0&\exp\left(-i\frac{\pi}{4}\right)
	\end{matrix}\right)=\exp(-i\frac{\pi}{8})R_{\mathrm{Z}}\left(-\frac{\pi}{4}\right)
\end{equation}
\section{Quantum circuits for preparing the $N=6$ DFS bases}
\label{ap6}
In this section the quantum circuits for preparing the $N=6$ decoherence free subspace bases are presented. The complete set of bases are taken to be $\{|a_1\rangle,|a_2\rangle,|a_3\rangle,|a_4\rangle,|a_5\rangle\}$, where
\begin{align}
	|a_1\rangle &= |\psi_{12}\rangle \otimes|\psi_{34}\rangle\otimes|\psi_{56}\rangle \nonumber\\
	|a_2\rangle &= |\psi_{12}\rangle \otimes|\psi_{35}\rangle\otimes|\psi_{46}\rangle \nonumber\\
	|a_3\rangle &= |\psi_{13}\rangle \otimes|\psi_{24}\rangle\otimes|\psi_{56}\rangle \nonumber\\
	|a_4\rangle &= |\psi_{13}\rangle \otimes|\psi_{25}\rangle\otimes|\psi_{46}\rangle \nonumber\\
	|a_5\rangle &= |\psi_{14}\rangle \otimes|\psi_{25}\rangle\otimes|\psi_{36}\rangle.
\end{align}
We explictly show the quantum circuits which are built based on Fig.~\ref{fig1}. The system qubits are labelled from $q_0$ to $q_5$, while the ancillary qubits which need to be measured are labelled with $q_6$, $q_7$, $\dots$.

\begin{figure*}[h]
	\centering
	\subfigure {\
		\begin{minipage}[h]{0.08\linewidth}
			\centering
			\begin{overpic}[scale=0.28]{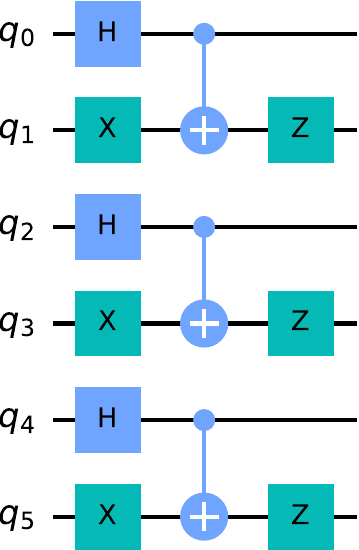}
				\put(-10,103){\textbf{(a)}}
			\end{overpic}
		\end{minipage}
	}
\quad
	\subfigure {\
		\begin{minipage}[h]{0.8\linewidth}
			\centering
			\begin{overpic}[scale=0.2]{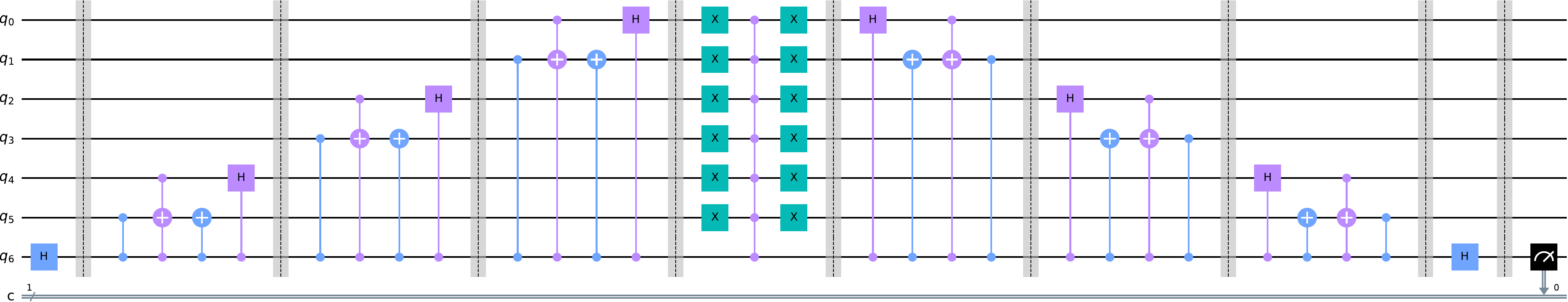}
				\put(0,20.15){\textbf{(b)}}
			\end{overpic}
		\end{minipage}
	}
	\caption{(a) The quantum circuit for preparing $|u_1\rangle$. (b) The quantum circuit for preparing $|u_2\rangle$.}
\end{figure*}
\begin{figure*}[h]
	\centering
	\includegraphics[width=.88\linewidth]{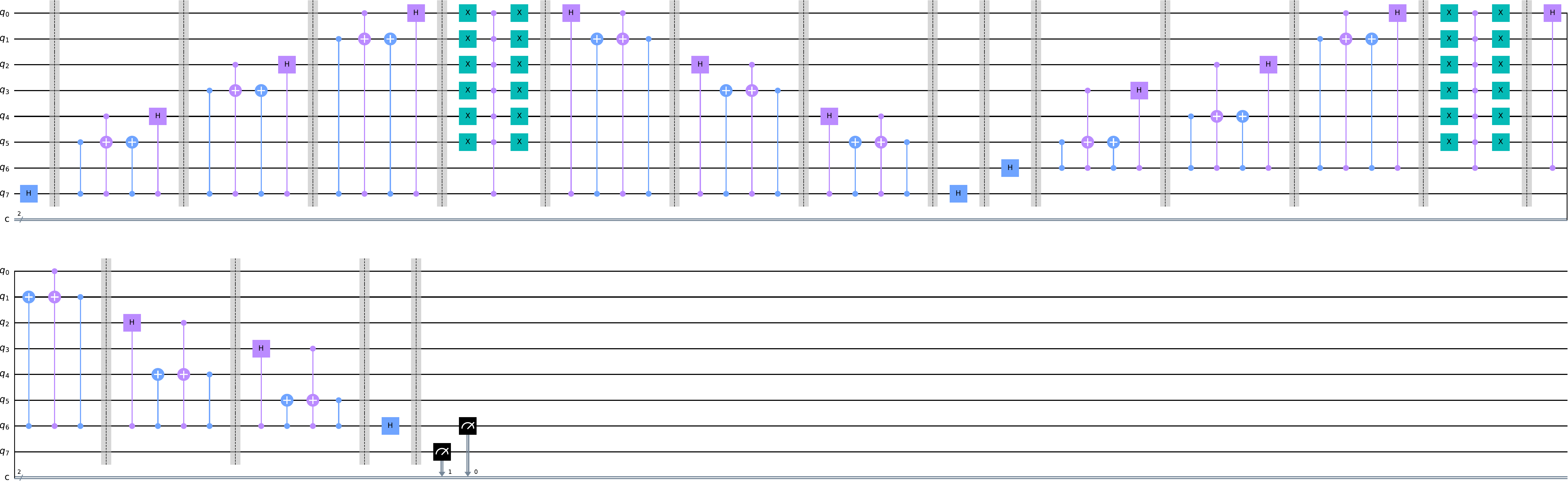}
	\caption{The quantum circuit for preparing $|u_3\rangle$}
\end{figure*}
\begin{figure*}[h]
	\centering
	\includegraphics[width=.88\linewidth]{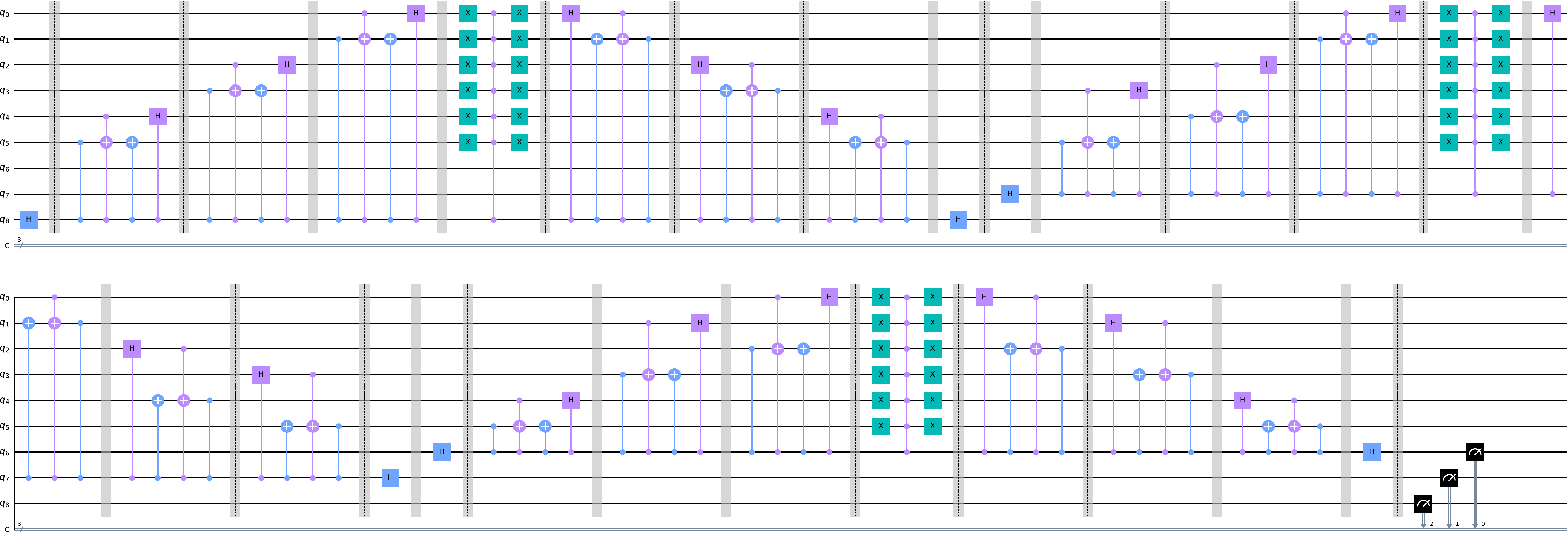}
	\caption{The quantum circuit for preparing $|u_4\rangle$}
\end{figure*}

\begin{figure*}[h]
	\centering
	\includegraphics[width=.88\linewidth]{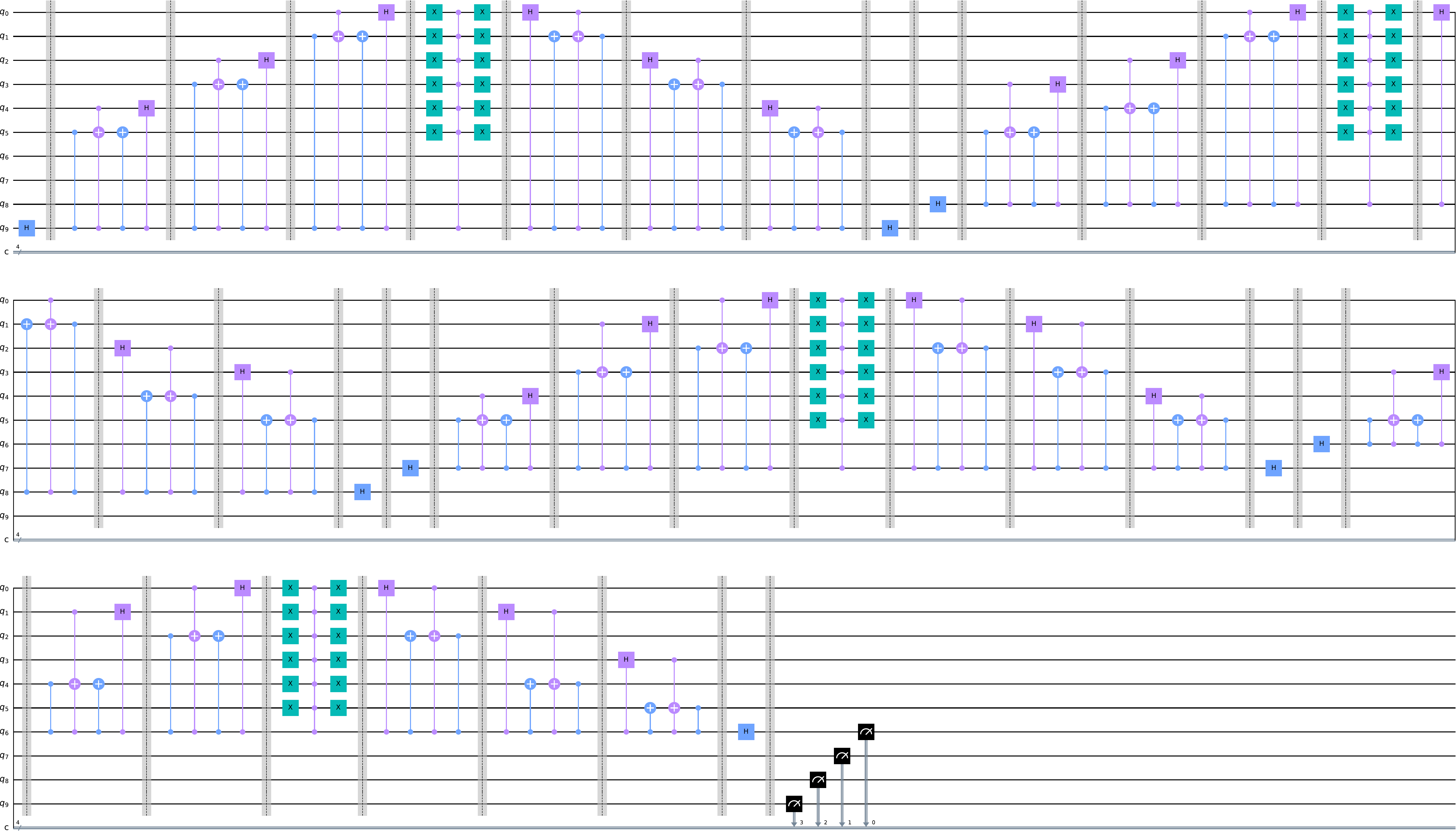}
	\caption{The quantum circuit for preparing $|u_5\rangle$.}
\end{figure*}


\end{document}